\documentclass[%
journal=jacsat,manuscript=article,
 amsmath,amssymb,
]{achemso}

\usepackage{graphicx}
\usepackage{dcolumn}
\usepackage{bm}
\usepackage[dvipsnames]{xcolor}
\usepackage{soul}
\usepackage[version=3]{mhchem}

\author{Joan M. Montes de Oca}\altaffiliation{Equal contribution}\affiliation[University of Chicago]{Pritzker School of Molecular Engineering, University of Chicago, Chicago, Illinois 60637, United States}
\alsoaffiliation[AMEWS]{Advanced Materials for Energy-Water Systems (AMEWS) Energy Frontier Research Center, Argonne National Laboratory, 9700 South Cass Avenue, Lemont, IL 60439, USA}

\author{Johnson Dhanasekaran}\altaffiliation{Equal contribution}\affiliation[University of Chicago]{Pritzker School of Molecular Engineering, University of Chicago, Chicago, Illinois 60637, United States} 
\alsoaffiliation[AMEWS]{Advanced Materials for Energy-Water Systems (AMEWS) Energy Frontier Research Center, Argonne National Laboratory, 9700 South Cass Avenue, Lemont, Illinois 60439, USA}

\author{Andr\'es C\'ordoba}\affiliation[University of Chicago]{Pritzker School of Molecular Engineering, University of Chicago, Chicago, Illinois 60637, United States}
\alsoaffiliation[AMEWS]{Advanced Materials for Energy-Water Systems (AMEWS) Energy Frontier Research Center, Argonne National Laboratory, 9700 South Cass Avenue, Lemont, Illinois 60439, USA}

\author{Seth Darling}\affiliation[University of Chicago]{Pritzker School of Molecular Engineering, University of Chicago, Chicago, Illinois 60637, United States}
\alsoaffiliation[Argonne1]{Chemical Sciences and Engineering Division, Argonne National Laboratory, Lemont, Illinois 60439, United States}
\alsoaffiliation[AMEWS]{Advanced Materials for Energy-Water Systems (AMEWS) Energy Frontier Research Center, Argonne National Laboratory, 9700 South Cass Avenue, Lemont, Illinois 60439, USA}

\author{Juan J. De Pablo}\affiliation[University of Chicago]{Pritzker School of Molecular Engineering, University of Chicago, Chicago, Illinois 60637, United States} \alsoaffiliation[Argonne2]{Materials Science Division, Argonne National Laboratory, Lemont, Illinois 60439, United States}
\alsoaffiliation[AMEWS]{Advanced Materials for Energy-Water Systems (AMEWS) Energy Frontier Research Center, Argonne National Laboratory, 9700 South Cass Avenue, Lemont, Illinois 60439, USA}
\email{depablo@uchicago.edu}

\title{Ionic transport in electrostatic Janus Membranes. An explicit solvent molecular dynamic simulation.}

\keywords{Power generation, Janus membrane, nanofluidics, Ionic transport, Non-equilibrium molecular dynamics}

\begin{document}

%
%
%

\begin{abstract}

Janus --or two-sided, charged membranes offer promise as ionic current rectifiers. In such systems, pores consisting of two regions of opposite charge can be used to generate a current from a gradient in salinity. The efficiency of \textcolor{black}{nanoscale} Janus pores increases dramatically as their diameter becomes smaller. However, little is known about the underlying transport processes\textcolor{black}{, particularly under experimentally accessible conditions}. In this work, we examine the molecular basis for rectification in Janus nanopores using an applied electric field. Molecular simulations with explicit water and ions are used to examine the structure and dynamics of all molecular species in aqueous electrolyte solutions. \textcolor{black}{For several macroscopic observables, the results of such simulations are consistent with experimental observations on asymmetric membranes. Our analysis reveals a number of previously unknown features, including a pronounced local re-orientation of water molecules in the pores, and a segregation of ionic species that has not been anticipated by previously reported continuum analyses of Janus pores. Using these insights, a model is proposed for ionic current rectification in which electric leakage at pore entrance controls net transport.}

\end{abstract}

\maketitle

\section{Introduction}\label{sec:intro}
Recent efforts to understand the transport of aqueous electrolyte solutions in nanofluidic devices have led to rapid advances. \textcolor{black}{Some of these can be attributed to the availability of improved nanofabrication and characterization techniques, while others have resulted from innovative ideas for energy harvesting and storage\citep{igal2015review,bocquet2020nanofluidicsage}. }
The search for devices that maximize ionic selectivity and current rectification while minimizing electric resistance, has led to an array of ingenious designs that are often characterized by symmetry breaking in the axial direction of the pore, and by having radial dimensions that are commensurate with the Debye length of the solution\cite{daiguji2005first-diode}. The so-called "Janus membranes", in particular, have garnered considerable attention due to their superior energetic efficiency and the ample design-space they offer \citep{sethdarling2018janus,zhang2017ultrathin,gao2014high,zhu2018JANUS-highsalinity}. \textcolor{black}{Janus membranes consist of two distinct regions with opposite charge. Such membranes exploit the asymmetry created between the two regions to establish unidirectional transport - a feature that is of interest for osmotic energy harvesting in situations where a strong built-in salinity gradient is available (e.g. rivers or oceans).} Past reports related to their performance have sometimes been accompanied by theoretical interpretations that rely on solution of the Poisson–Boltzmann (PB) or Poisson–Nernst–Planck (PNP) equations\citep{constantin2007poisson,wen2019ICR,cheng2010diodes-REVIEW,gao2014high,zhang2017ultrathin}. \textcolor{black}{As helpful as continuum approaches have been for design of Janus pores, it is unclear whether such models are capable of incorporating all of the relevant physics that arise in nanometer pores \cite{bazant2019GAPS-singledigit,bocquet2010review-SCALES}.} Familiar continuum concepts, such as that of a dielectric constant, viscosity, and diffusion coefficient must be revisited in pores whose dimensions are comparable to those of the range of interaction between ionic species. 

\textcolor{black}{The primary goal of this work is to understand the molecular origin of current rectification in Janus membranes. While this question has been addressed using continuum representations\citep{gao2014high,zhang2017ultrathin}, molecular dynamics simulations of rectification have been scarce\citep{cruz2009first-ICR-MD,hato2017implicit-solvent-MD,reviewers_paper}. In particular, all-atom studies have focused on much smaller nanopores than those used in experiments, and have not considered the complex geometries that have been shown to enhance rectification in experiments. The nanopore model considered here is inspired by recent experimental reports, and it exhibits the main features of fully-functional nanoscale laboratory devices reported in the literature\citep{gao2014high,zhang2017ultrathin}. In the confinement regime considered here (5-10 nm pore diameter), the three-dimensional hydrogen bond (HB) network of the solution is perturbed but not destroyed. At the same time, electrostatic effects are not too strong because the smallest geometric scale is much larger than the Debye length of the 0.05M electrolyte solutions in our simulations (1.36 nm). Furthermore, when the concentration is raised to 0.17M, that Debye length drops to 0.73 nm. This regime is of interest because electrostatic interactions play a key role in the underlying physical processes, but they do not dominate the pairwise interaction energy between molecules, thereby leading to opportunities for more effective control of ionic transport.}    

By \textcolor{black}{relying on large-scale} molecular dynamic simulations, we are able to correctly reproduce many of the experimentally observed membrane characteristics, including the current-voltage response, the charge density distribution, and the ionic current rectification (ICR) factor. Having established the ability of the pore model adopted here to reproduce key experimental features, we are able to identify the underlying molecular processes responsible for the observed behaviour.

\textcolor{black}{Our results suggest that ICR in Janus membranes is linked to the ability of the electric field induced inside the membrane to pull in ions from the reservoir. The asymmetry of the membrane is such that, when a negative bias is applied, the induced field is not able to reach the ions in the reservoir, thereby inhibiting ionic transport. Beyond the asymmetrical ionic distribution} inside the pore in response to an applied bias, we find that water also reacts asymmetrically. Water dipoles align strongly with the electric field produced by the charged walls, and also with the externally applied field. Importantly, the dipole alignment is found to be strongly anisotropic, and to depend on the local ionic concentration. \textcolor{black}{The new insights provided by the simulations presented here should be of considerable assistance for development of improved continuum theories for description of electrolytes in nanopores, and for design of new devices with better ICR characteristics.}

\section{Results and discussion}\label{sec:results}

\begin{table}
\caption{\textcolor{black}{Ionic current generated in the configurations considered in this study, including two concentrations ($C$) and three applied voltages. The rectification factor $R$ is calculated by dividing the current obtained for opposite external biases. The current enhancement is defined as the ratio between the current obtained for the two concentrations at the same bias.}}
\begin{tabular}{ccccc}
\hline
Concentration [M] & I [nA] activated & I [nA] 0V & I [nA] inactivated & $R$ ($I_+/I_-$)\\
\hline
0.05 & 3.037 & 0.022 & -0.341 & 8.91\\
\textcolor{black}{0.17} & \textcolor{black}{6.079} & \textcolor{black}{0.048} & \textcolor{black}{-1.858} & \textcolor{black}{3.27}\\
\hline 
\textcolor{black}{I enhancement} & \textcolor{black}{2.00} & & \textcolor{black}{5.45} &\\
\hline
\end{tabular}
\label{table:rectification}
\end{table}

We start our discussion with a general characterization of the Janus membrane, and then examine the molecular mechanisms that influence it's behaviour. 
Perhaps the most important indicator of performance is the ionic current that is generated in response to an applied potential, shown in Table \ref{table:rectification}. Note that a non-linear response to the applied potential is desirable, and is the reason for the asymmetrical design of the membrane \cite{cheng2010diodes-REVIEW,igal2015review,siwy2006-ICR-review,sethdarling2018janus}.  
A qualitative comparison of our results with those reported in the literature \cite{gao2014high,zhang2017ultrathin,zhu2018JANus-rectification_factor2} reveals that all the major features observed in experiments on Janus membranes are reproduced by our molecular model, including the ionic selectivity for sodium over chloride, and a significant rectification factor, defined as the ratio between the ionic current of the active state and the inactive state.

\textcolor{black}{The current rectification factor is a crucial indicator of the membrane's performance.} It is also independent of the number of pores, thereby allowing for direct comparison between our results and experimental measurements. Our simulations show a rectification factor of \textcolor{black}{8.91 at a 0.05M electrolyte concentration, and 3.27 at a 0.17M concentration, which are consistent with the experimental value of 7.1 reported by Zhang \latin{et al.} for a similar electrolyte concentration \cite{zhang2017ultrathin}. Other studies on Janus membranes of similar design have reported rectification values of comparable magnitude \cite{zhu2018JANus-rectification_factor2,gao2014high}.} \textcolor{black}{In agreement with experiments, our simulations show a drop in rectification with increasing electrolyte concentration, as evidenced by the 37\% reduction reported in Table \ref{table:rectification} when the reservoir concentration is changed from 0.05M to 0.17M.} 


\begin{figure*}
\hspace*{-0cm}
\includegraphics[scale=0.5]{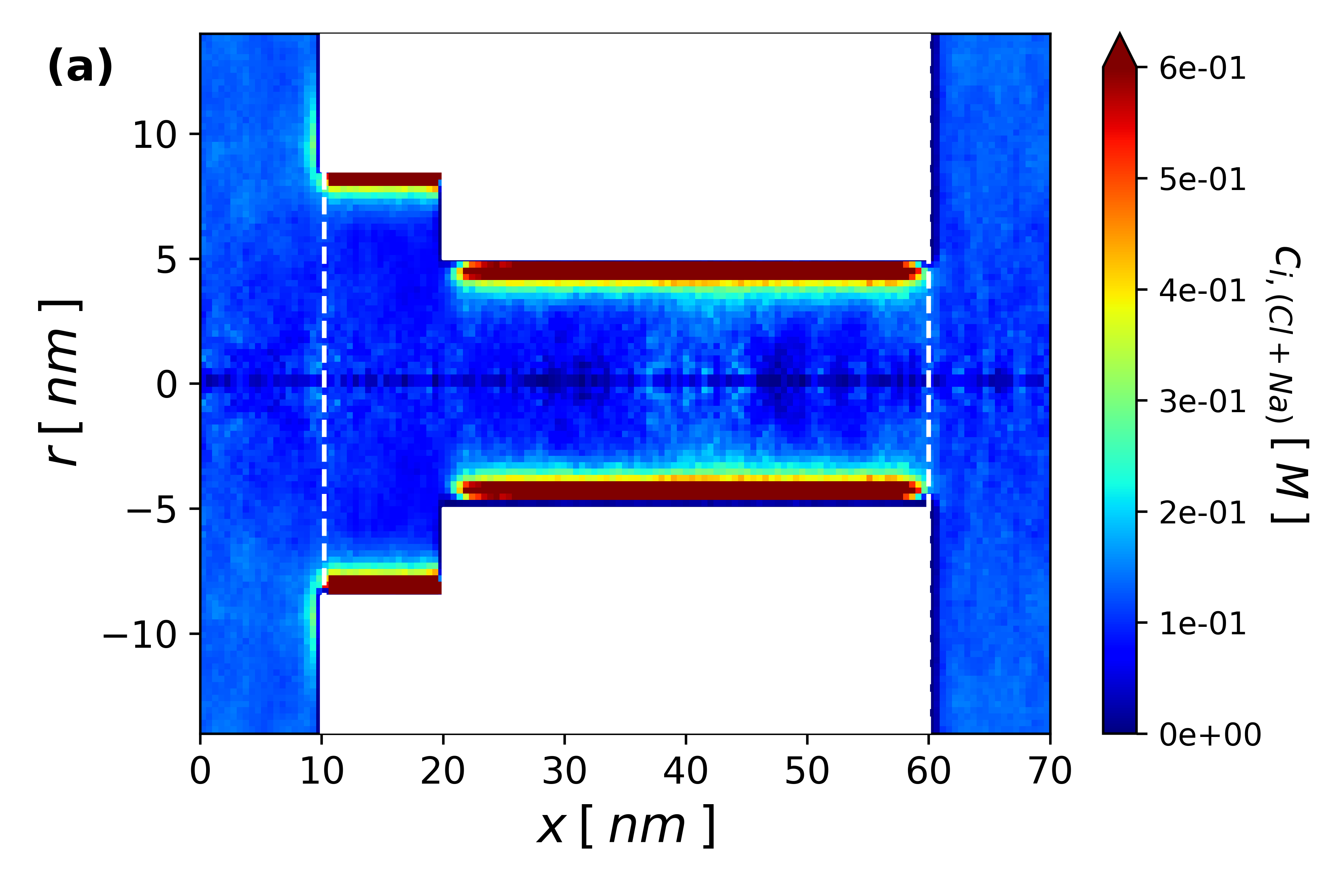}
\includegraphics[scale=0.5]{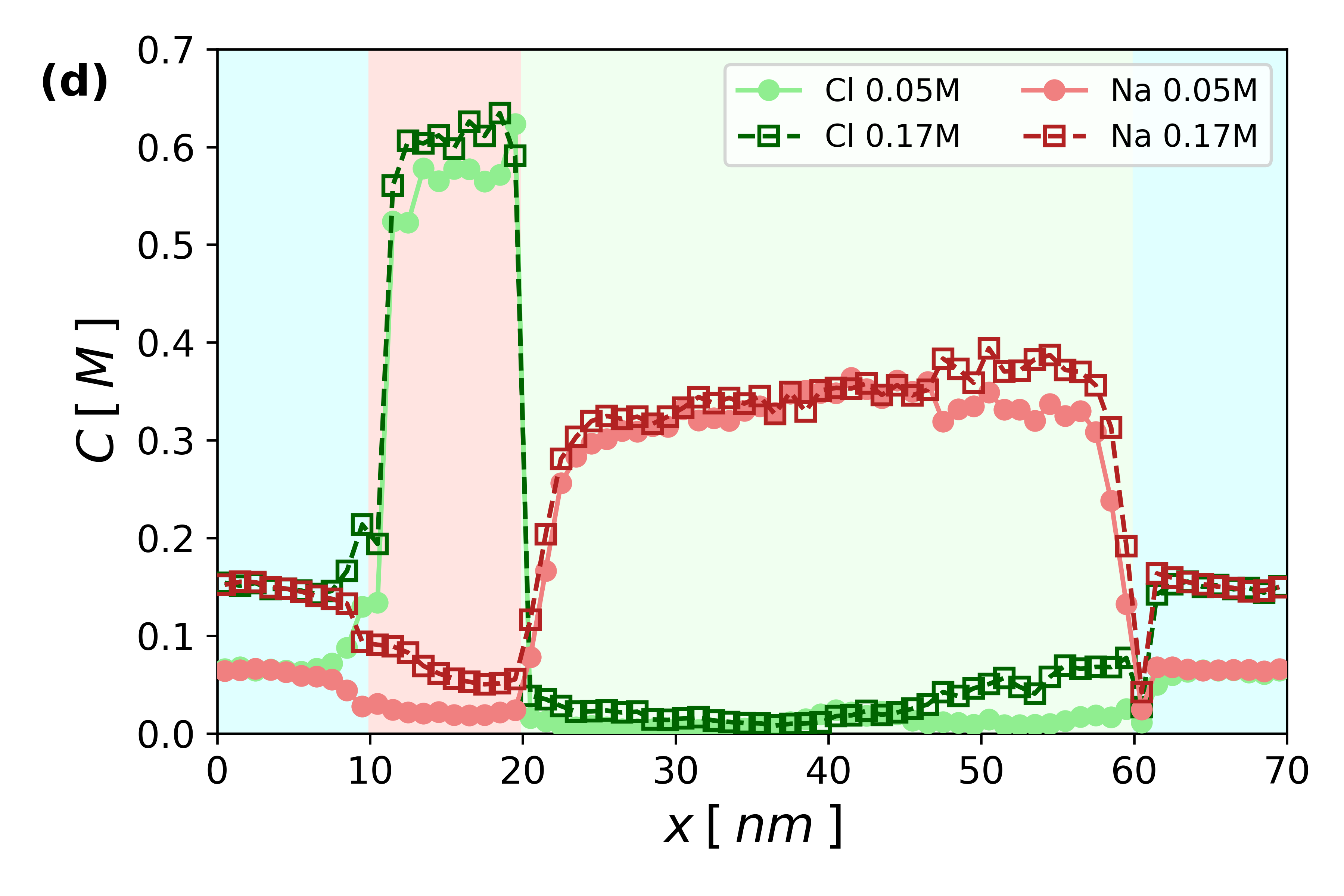}
\includegraphics[scale=0.5]{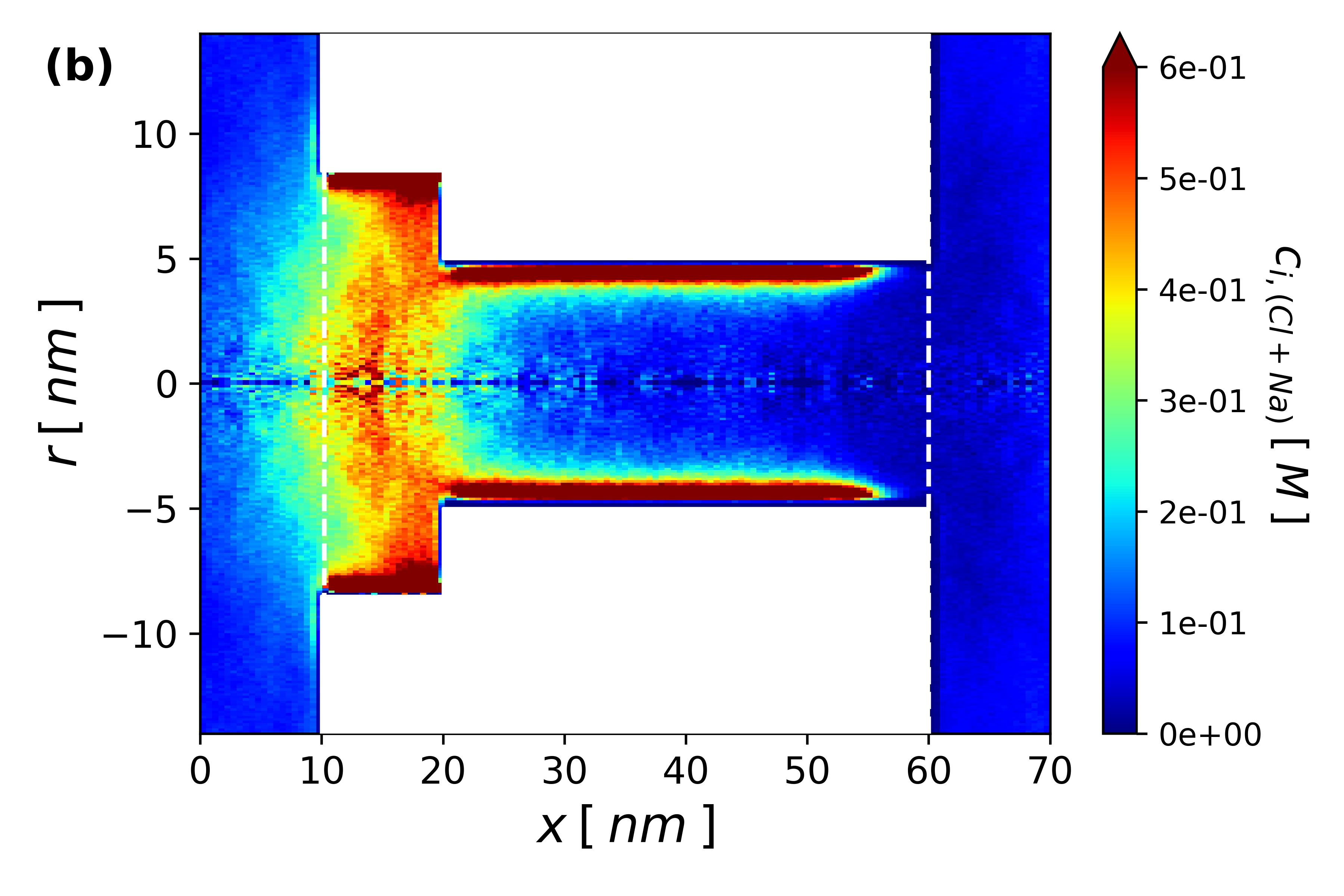}
\includegraphics[scale=0.5]{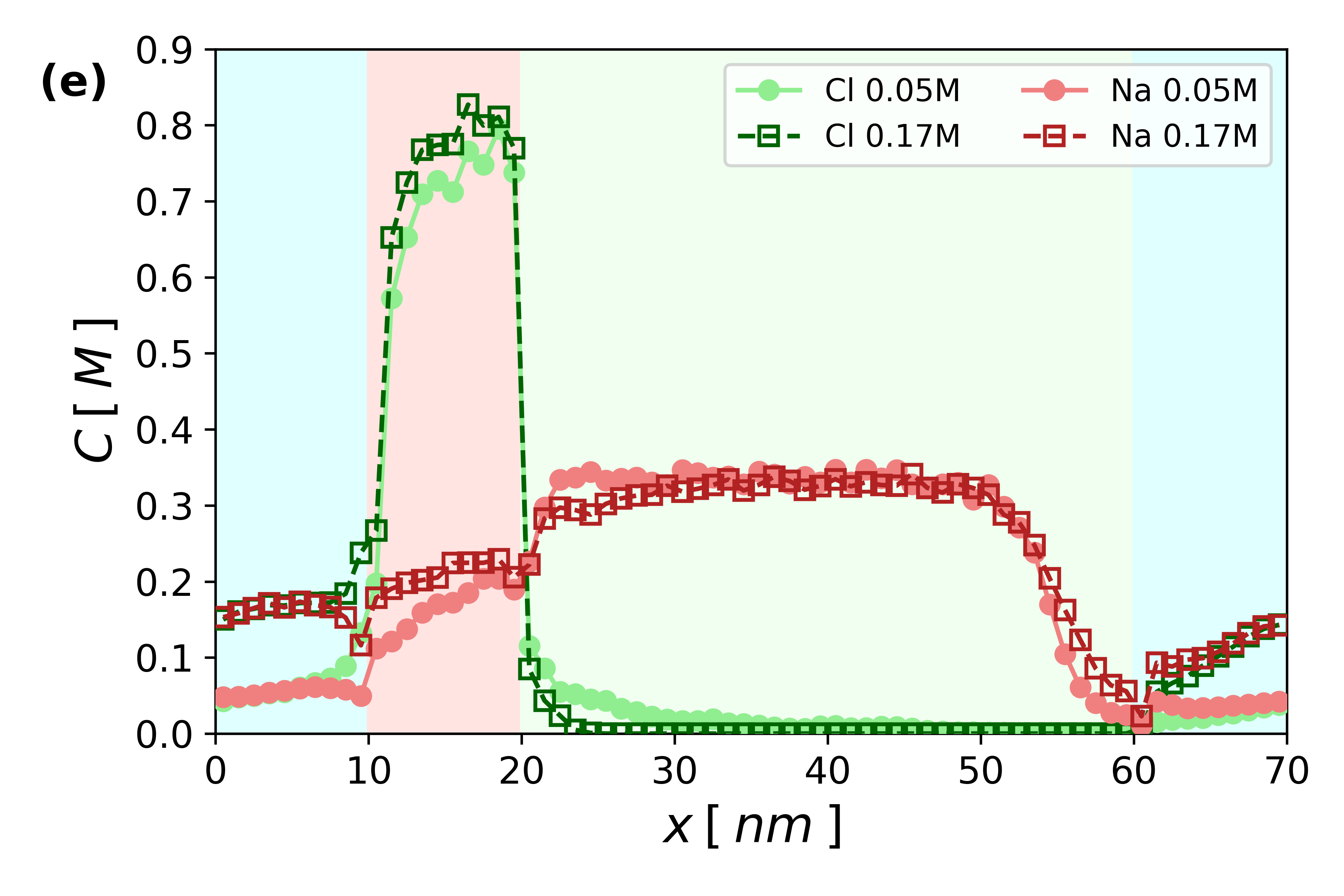}
\includegraphics[scale=0.5]{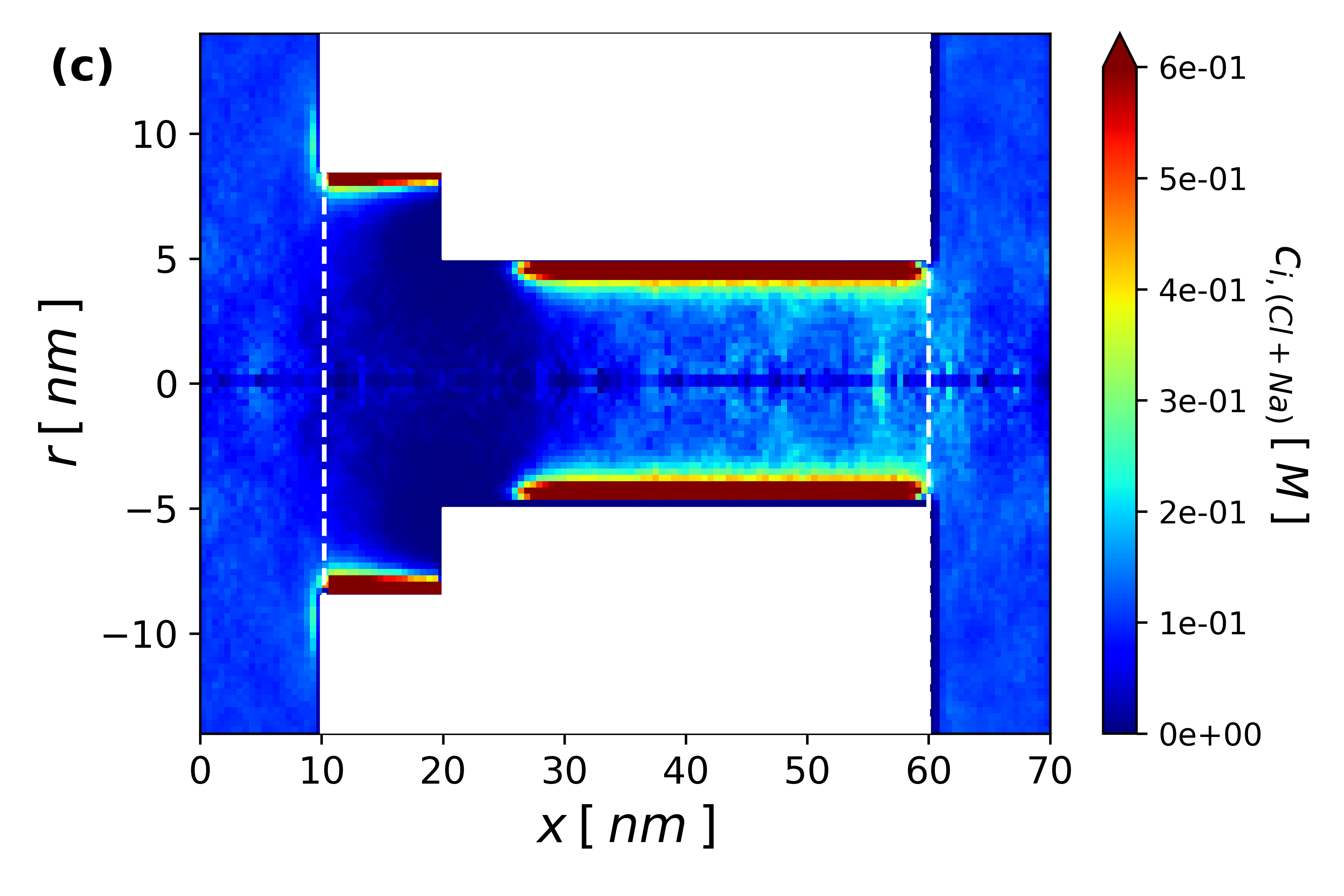}
\includegraphics[scale=0.5]{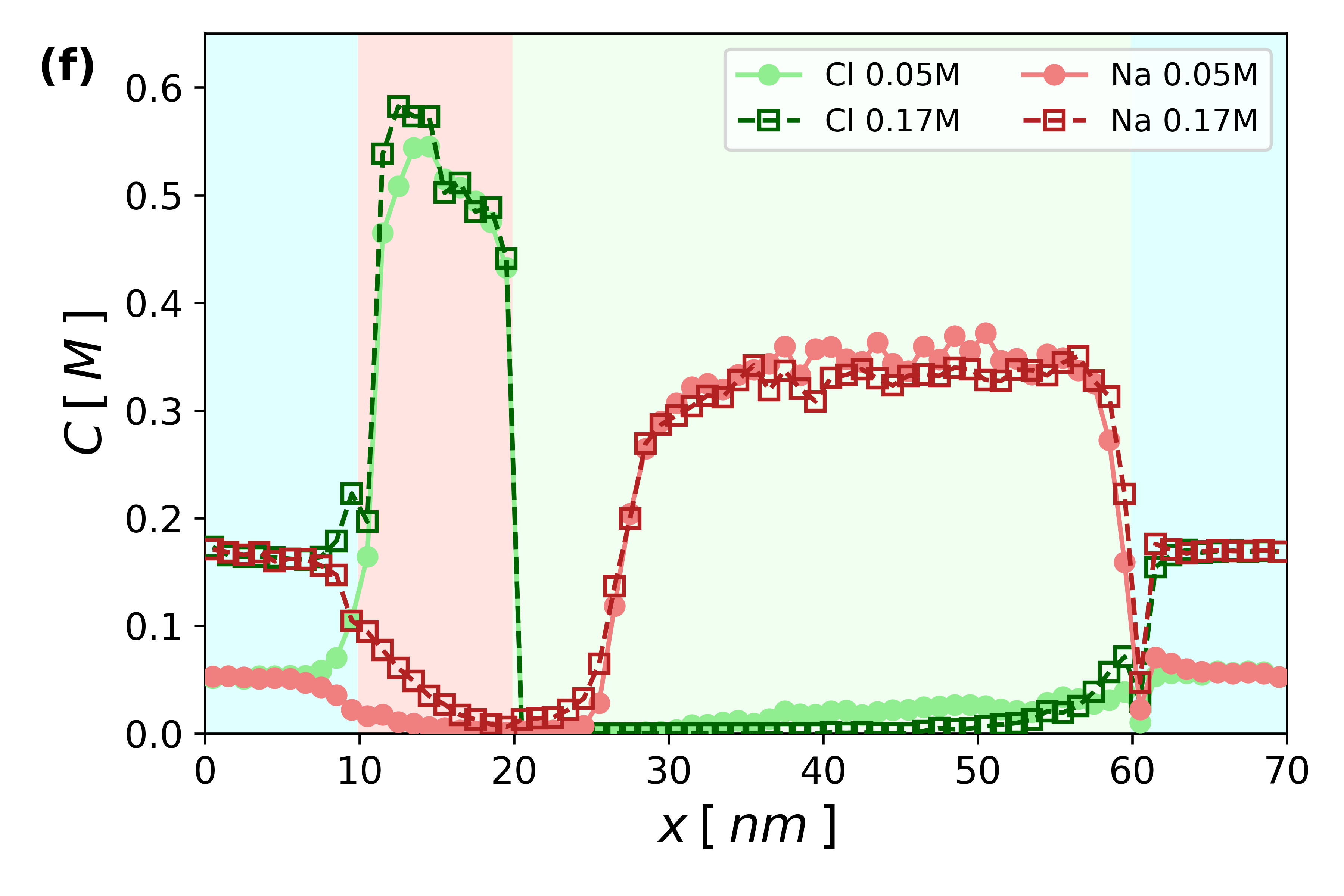}
\caption{2D Ionic (Na+Cl) concentration distribution \textcolor{black}{(left column) and the radially averaged concentration individual sodium and chloride projected onto $x$ (right column) are shown. The top row is the equilibrium state at 0V, the middle row is the activated state while the bottom row is inactivated case. The 2D ion concentration plot is for a reservoir concentration of 0.05 M. For the higher 0.17 M case please see the supporting information Figure S 2.}}
\label{fig:concentration}
\end{figure*}

We next consider the average distribution of ions in the absence and presence of an \textcolor{black}{applied electric field.} The mechanism for ionic current rectification is generally understood in terms of the ion accumulation and depletion model initially proposed by Woermann for conical pores \cite{woermann2003ICR-original}, and later extended to generic asymmetrical channels \cite{siwy2006-ICR-review,cheng2010diodes-REVIEW,white2016ICR,wen2019ICR,boquet2019beyondtradeoff}. Figure \ref{fig:concentration} left column shows the ionic concentration profiles along the radial and axial coordinates for both the active \textcolor{black}{(applied electric field 0.0333 V/nm)} and inactive versions of the pore \textcolor{black}{(external electric field -0.0333 V/nm)}, along with the unbiased equilibrium state. Consistent with the current understanding of current rectification, we find the inactive state to exhibit a reduced ionic content with respect to the active state\textcolor{black}{\cite{gao2014high,zhang2017ultrathin,zhu2018JANUS-highsalinity,reviewers_paper}}. However, in contrast to the Woermann model of ICR, we find that the pore is never fully depleted of ions with respect to the bulk. Instead, the concentration of ions inside the channel is always greater than that in the reservoir.

\begin{figure}
\hspace*{-0cm}
\includegraphics[scale=0.75]{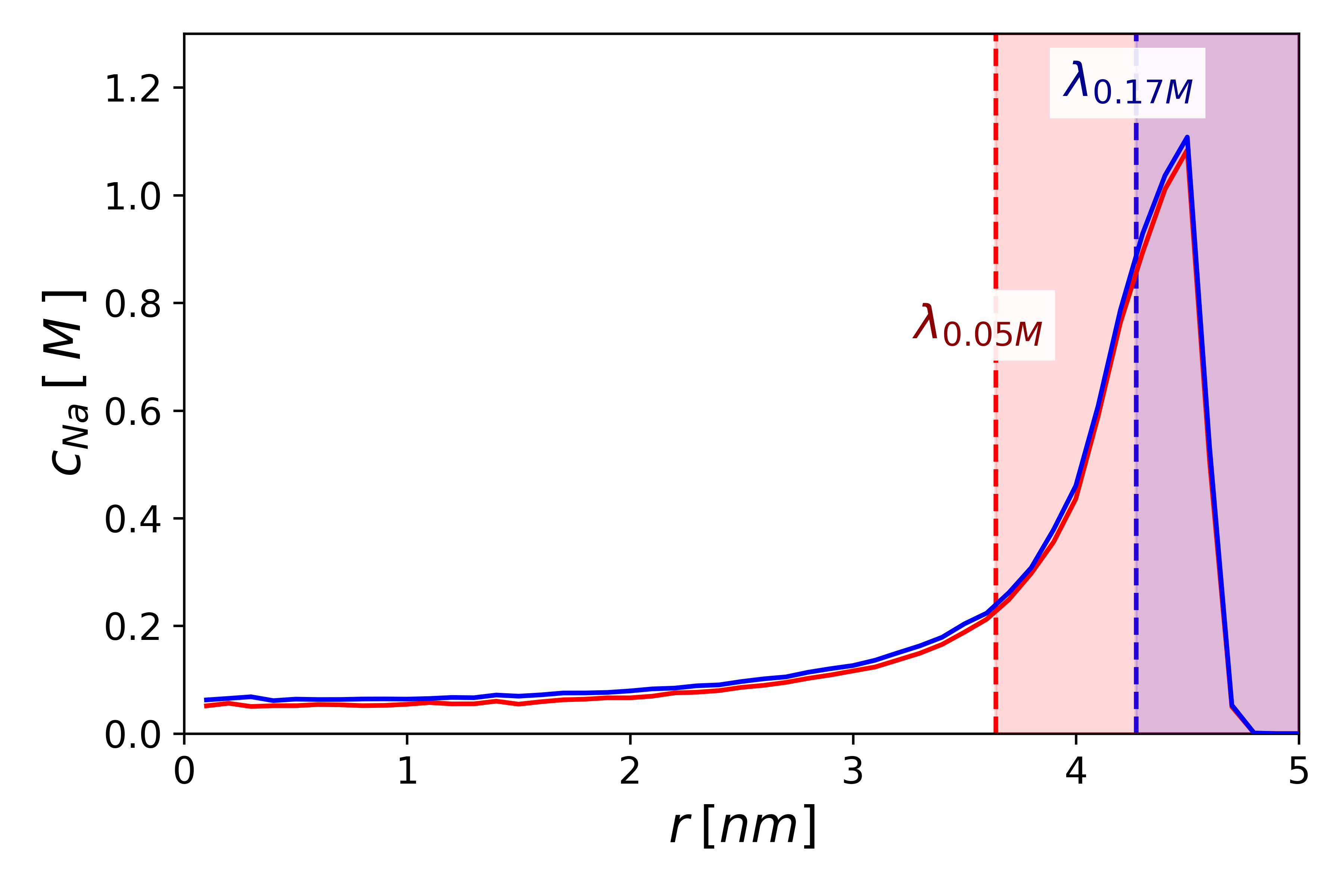}
\caption{\textcolor{black}{Radial concentration distribution of sodium ions in the narrow channel, obtained by averaging over its middle portion $30~{\rm nm} < x < 50~{\rm nm}$, as a function of $r$. The results for a reservoir salt concentration of 0.05 M and 0.17 M are nearly identical, even though the two Debye lengths, shown using dashed lines, are different.}}
\label{fig:radial_conc_plot}
\end{figure}

\textcolor{black}{In Figure \ref{fig:concentration}, the three right-hand column plots show a comparison of the ionic distribution inside the pore for the two electrolyte solution concentrations considered here. For clarity, the ionic concentration has been projected only over the axial coordinate, and split into the anion and cation contributions. The addition of ions does not translate into a substantial change in the distribution of ions within the pore. The bulk of the extra ions is accommodated by the reservoir, raising the total concentration in that region from about 0.05M to 0.17M. This implies that the screening of the charged walls of the pores is barely altered. Figure \ref{fig:radial_conc_plot} compares the sodium radial concentration profiles in the narrow pore for the two reservoir concentrations. Both configurations exhibit a similar profile, even in the presence of very different Debye electric double layer thicknesses.}

\begin{figure}
\hspace*{-1cm}
\includegraphics[scale=0.7]{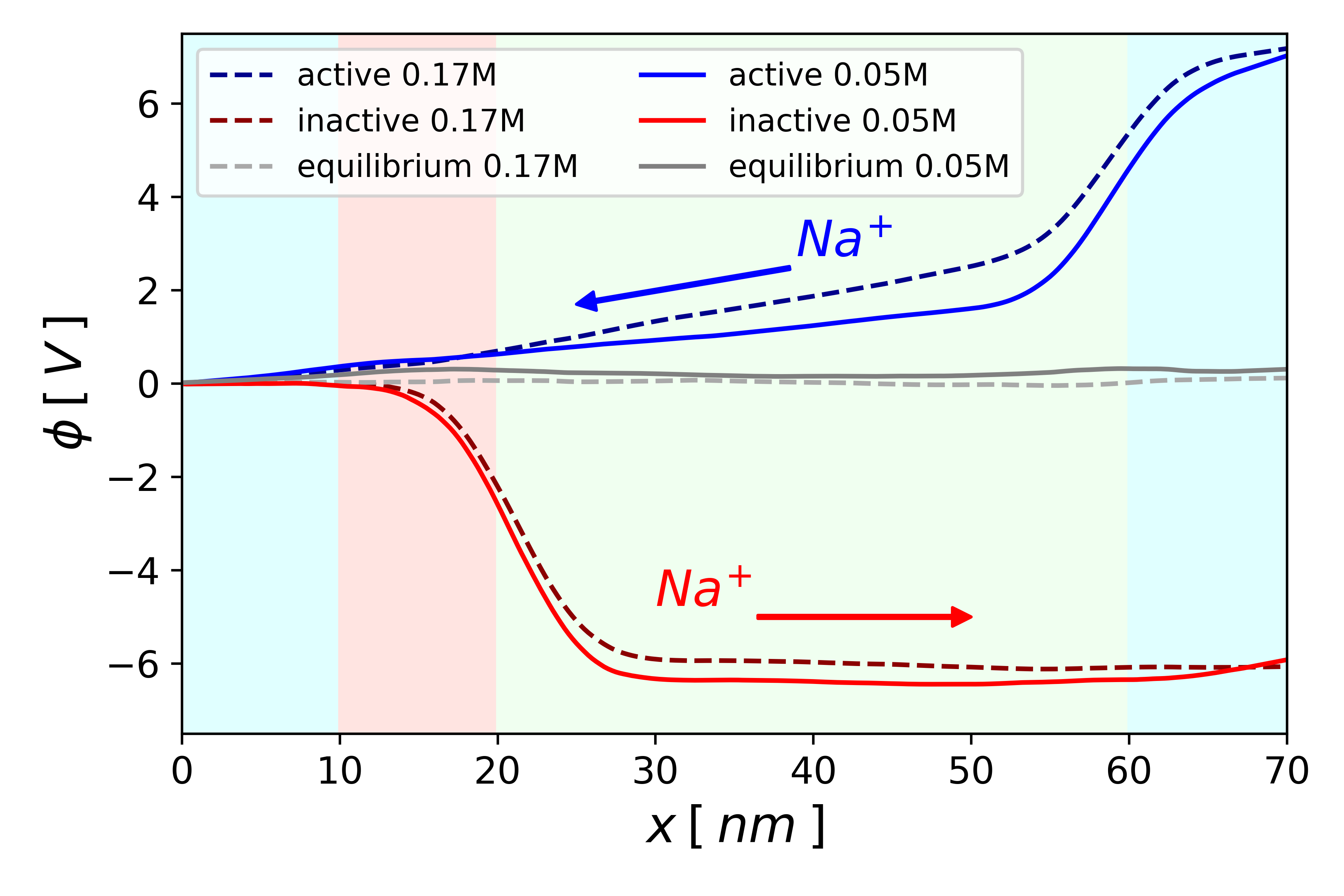}
\caption{\textcolor{black}{$\phi$ averaged over the radial coordinate, as a function of $x$. We show six cases: two reservoir electrolyte concentrations, in the activated, inactivated and equilibrium states. Also shown is an illustration of the direction taken by the dominant current carrying sodium ions. $\phi$ has been computed with a reference of 0 at $x=0$ and integrating the electric field reported in \ref{fig:electric_field}.}}
\label{fig:potential}
\end{figure}
\begin{figure*}
\hspace*{-0cm}
 \includegraphics[scale=0.7]{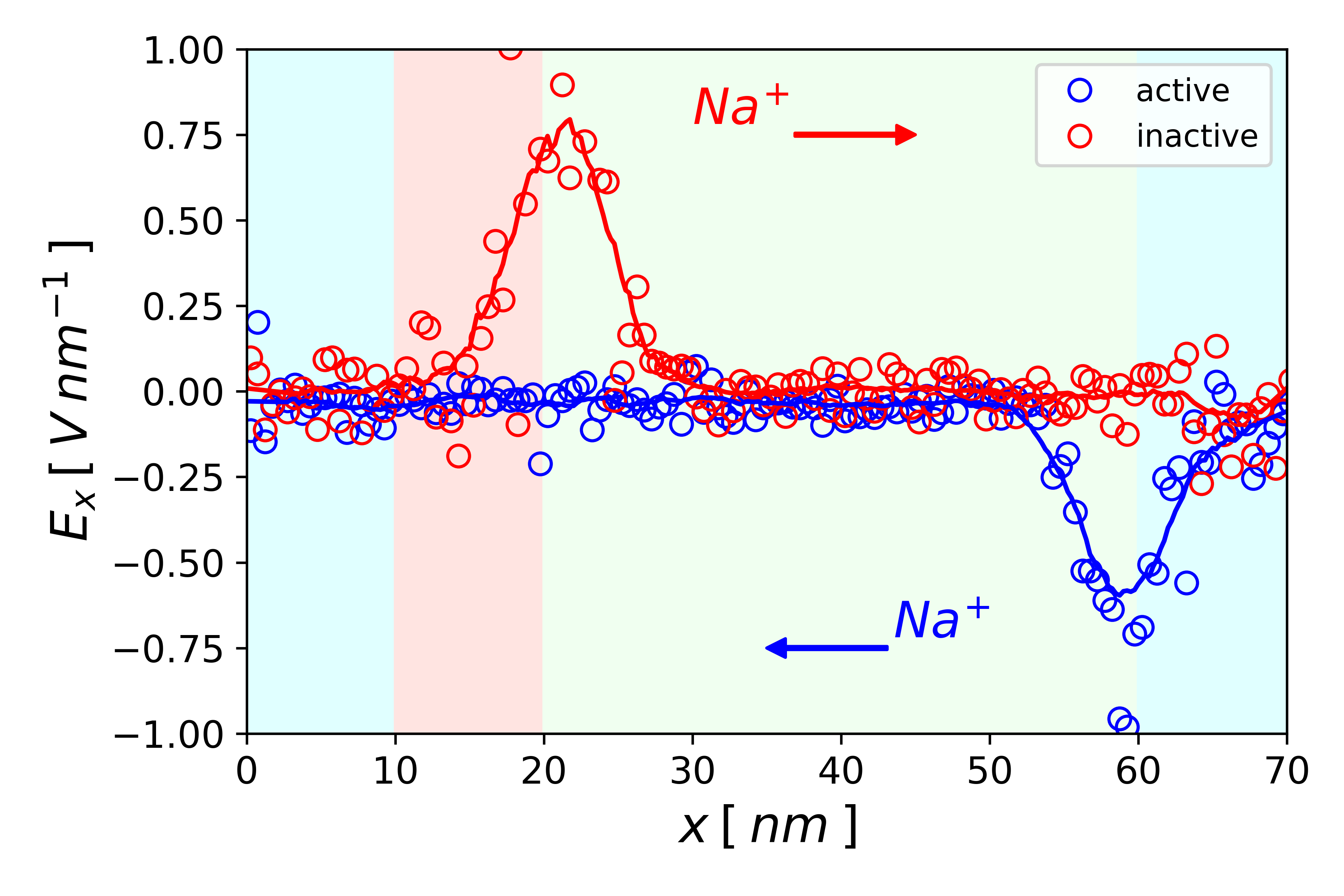}
\caption{\textcolor{black}{The radially averaged local electric field along the $x$ direction is shown as a function of the axial coordinate $x$ for the active and the inactivate states. At equilibrium, the electric field can be barely detected, and is not included here. For reference, the direction of sodium ion transport is also included. The curves shown here are for the reservoir concentration of 0.05M. For the 0.17M result, readers are referred to the supplementary Figure S3. The electric field is calculated from the electrostatic force generated on the ions. This is converted to an electric field via knowledge of the charge. The spatial ($r$, $x$) electric field is obtained by averaging over multiple ion instances crossing that location.}}
\label{fig:electric_field}
\end{figure*}

\textcolor{black}{While most of the pore interior is unaffected by the higher electrolyte concentration, there is a noticeable uptick in the local ion concentrations near the sharp edges at the entrance of the narrow pore, which exposes the wall charge that would have been screened in the equilibrium state. This leads to a significant drop in $\phi$ around a value of $x$ of 60 nm for the active state when the cations approach from the right, and around 20 nm for the cations moving from left to right in the inactive pore. This can be appreciated in the radially averaged local electric ($\phi$) in Figure \ref{fig:potential}. It shows a symmetric $\phi$ profile when adjusted for the direction of ionic movement. These results also confirm that sodium ions are the dominant charge carriers, even if the net charge of the lumen is positive. To show both cation dominance and symmetry more clearly, we also present the radially averaged electric field along the $x$ direction ($E_x$) in Figure \ref{fig:electric_field}. The peak location and width are set by the sodium ions for both non-equilibrium states of the channel.}

\begin{figure*}
\hspace*{-0cm}
 \includegraphics[scale=0.65]{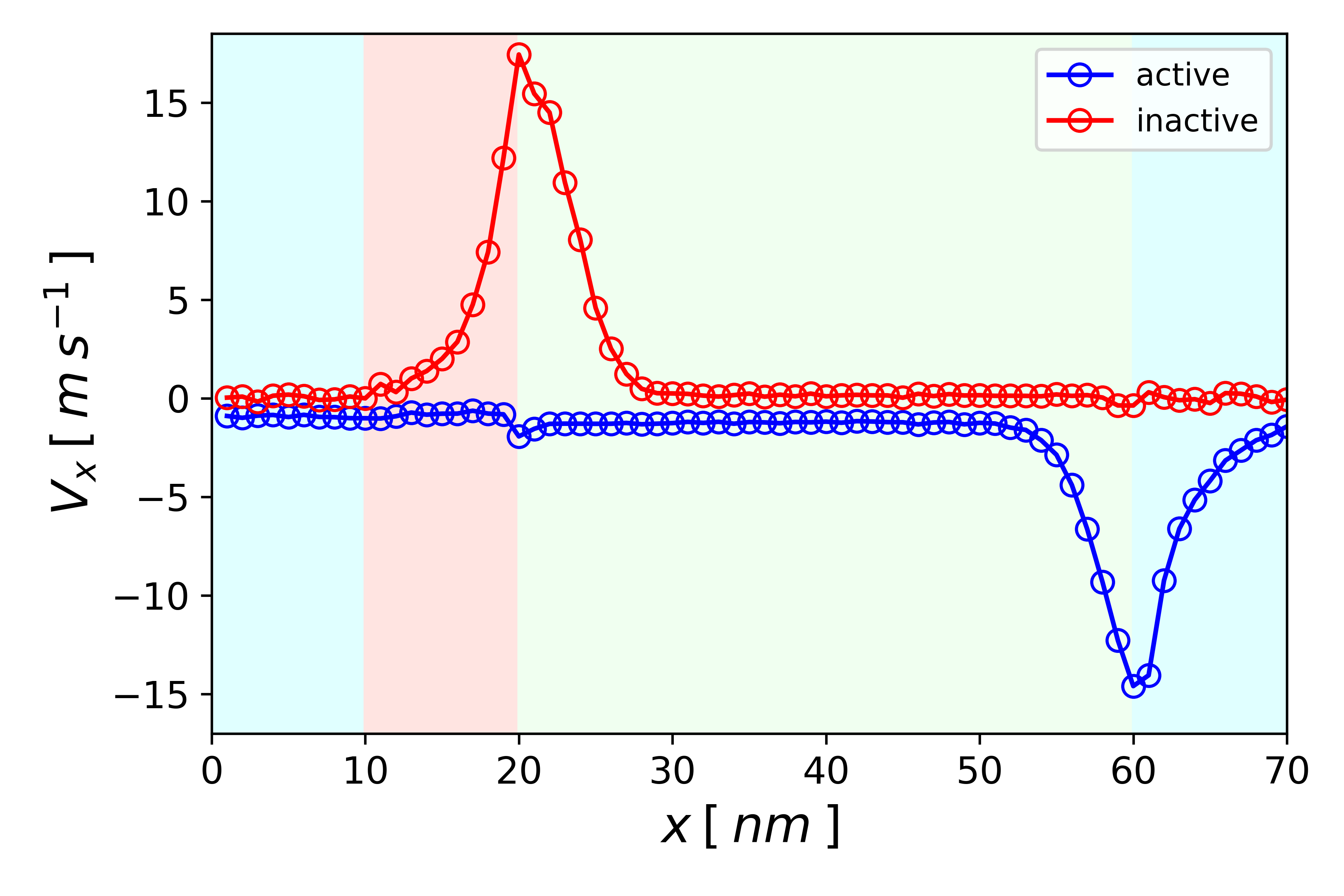}
\caption{\textcolor{black}{Sodium axial velocity, averaged over the radial coordinate, as a function of position $x$ for both the active and inactive states of the channel. The ions in the reservoir and in contact with the outer wall are excluded from the average. A complete 2D velocity snapshot is presented in Figure S 4 of the supporting information.}}
\label{fig:velocity}
\end{figure*}
\textcolor{black}{There is no asymmetry in the underlying field; the rectification is driven by the local concentration changes that arise from the electrical double layer near the entrance of the narrow section of the pore. Measuring these local concentrations reveals a two-fold rise in the active case and a five-fold increase in the inactive state when the reservoir concentration is increased from 0.05M to 0.17M. These increases are directly correlated with the higher ionic current reported in Table \ref{table:rectification}. To understand ICR further, we also consider the possibility that the asymmetry between the active and inactive modes of the pore might be related to diffusion. To that end, we report the radially averaged axial sodium velocities ($V_x$) along $x$ in Figure 5. We once again recover the symmetry, and the qualitative features of the sodium velocity are consistent with the electric field reported in Figure \ref{fig:electric_field}, underscoring that enhanced diffusion does not account for the magnitude of $R$ of nearly 9 reported in Table \ref{table:rectification}.}

\textcolor{black}{Our proposed electric-field leakage model for ICR is based on both the electric force and the concomitant acceleration that starts well before the ion-depleted regions of the small section of the pore. In the active pore, ions are brought to the entrance from as far as 5 nm away, in the reservoir, by the electric field generated by the pore when voltage is applied. Similarly, in the inactive pore, ions are accelerated from remote positions but, this time, in the interior of the positively charged chamber, where the sodium concentration is much lower (even lower than in the bulk). Thus, the impact of ionic concentration on current is through an increase in the probability of crossing a barrier, and not necessarily a change in the barrier height.}

\begin{figure*}
\hspace*{-0cm}
\includegraphics[scale=0.25]{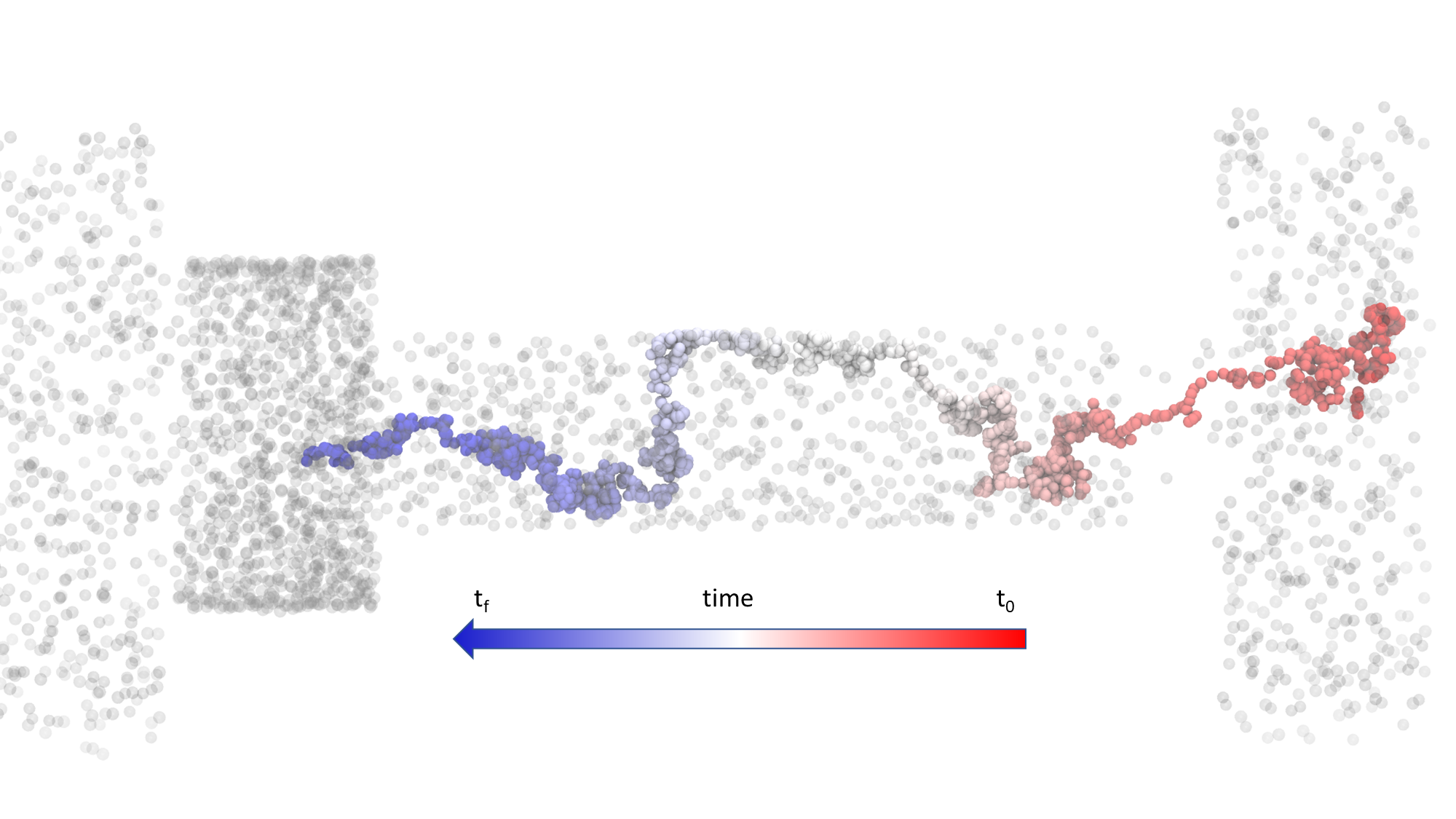}
\caption{\textcolor{black}{Representative trajectory of a sodium ion, the dominant charge carrier, across the membrane. It begins with classical Brownian motion in the reservoir. As it enters the narrow pore , the ion is driven to the wall and dragged along its surface. As the ion approaches the wider portion of the pore, it is forced to the center of the channel. The return to Brownian motion on the other side of the pore is not shown for conciseness.}}
\label{fig:acceleration}
\end{figure*}

\textcolor{black}{To better illustrate the dynamics of the Janus membrane operation we show in Figure \ref{fig:acceleration}the path taken by a sodium ion as it traverses the nanopore. We can observe a Brownian motion mechanism, aided by the applied electric field, through which the sodium ion approaches the membrane entrance. There, it spends a considerable amount of time before it crosses over to the negatively charged section of the pore. Once in that section, it is rapidly shuttled to the wall and dragged towards the wider section, where it is pushed to the center of the channel.}

\textcolor{black}{We now turn our focus towards} the explicit role of water in a system with dimensions comparable to those used in laboratory experiments. It is well known that water behaviour deviates from that of the bulk under extreme confinement conditions. In the context of nanofluidics, this level of confinement is often understood to occur below one nanometer, where the size of the system is comparable to molecular dimensions \cite{bazant2019GAPS-singledigit,kavokine2021netz-boquet-REVIEW,corti2021structure}. However, the size of the pore considered here is one order of magnitude above that scale, and therefore in a regime that has been assumed in the past to approach the continuum limit. In that regime, classical theories are expected to be valid, but have been less explored by MD simulations \cite{bocquet2010review-SCALES}. \textcolor{black}{In what follows, we only highlight some of the most salient features of the structure and dynamics of water in our simulated Janus membrane.}

One of the main reasons that water behaviour is difficult to predict is that molecular interactions propagate well beyond the size of a single molecule through a complex hydrogen bonding network. Furthermore, many of the unusual phenomena observed in nanofluidic devices occur in pore sizes that are too small to accommodate a fully three-dimensional HB network \cite{zhou2020subnanometer-barriers,hummer2001nanoclasic,alarcon2014hydrophobicity}. Our results indicate that the number of hydrogen bonds per molecule inside the pore is similar to the value computed in the reservoir.

\begin{figure}
\hspace*{-0cm}
\includegraphics[scale=0.65]{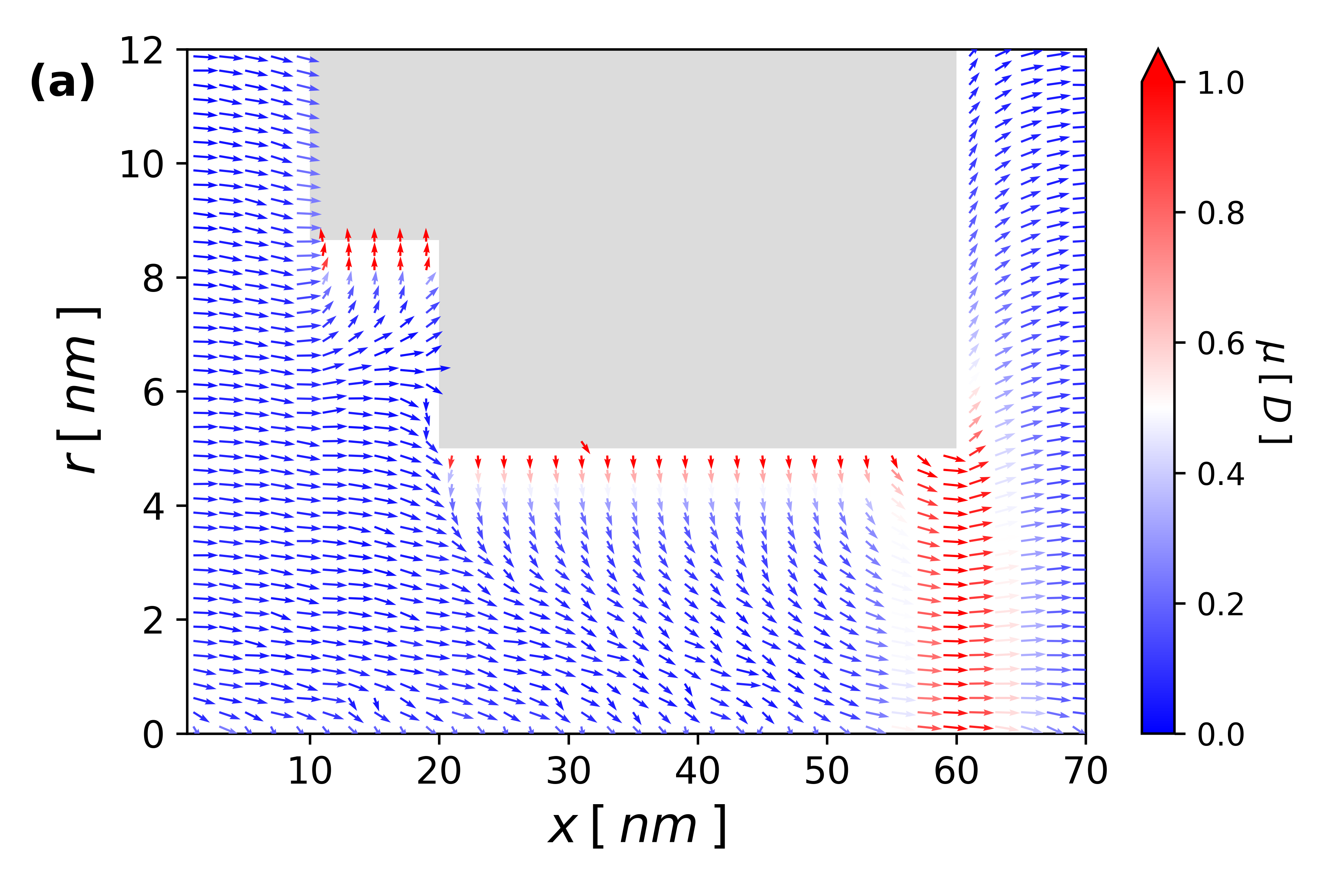}
\includegraphics[scale=0.65]{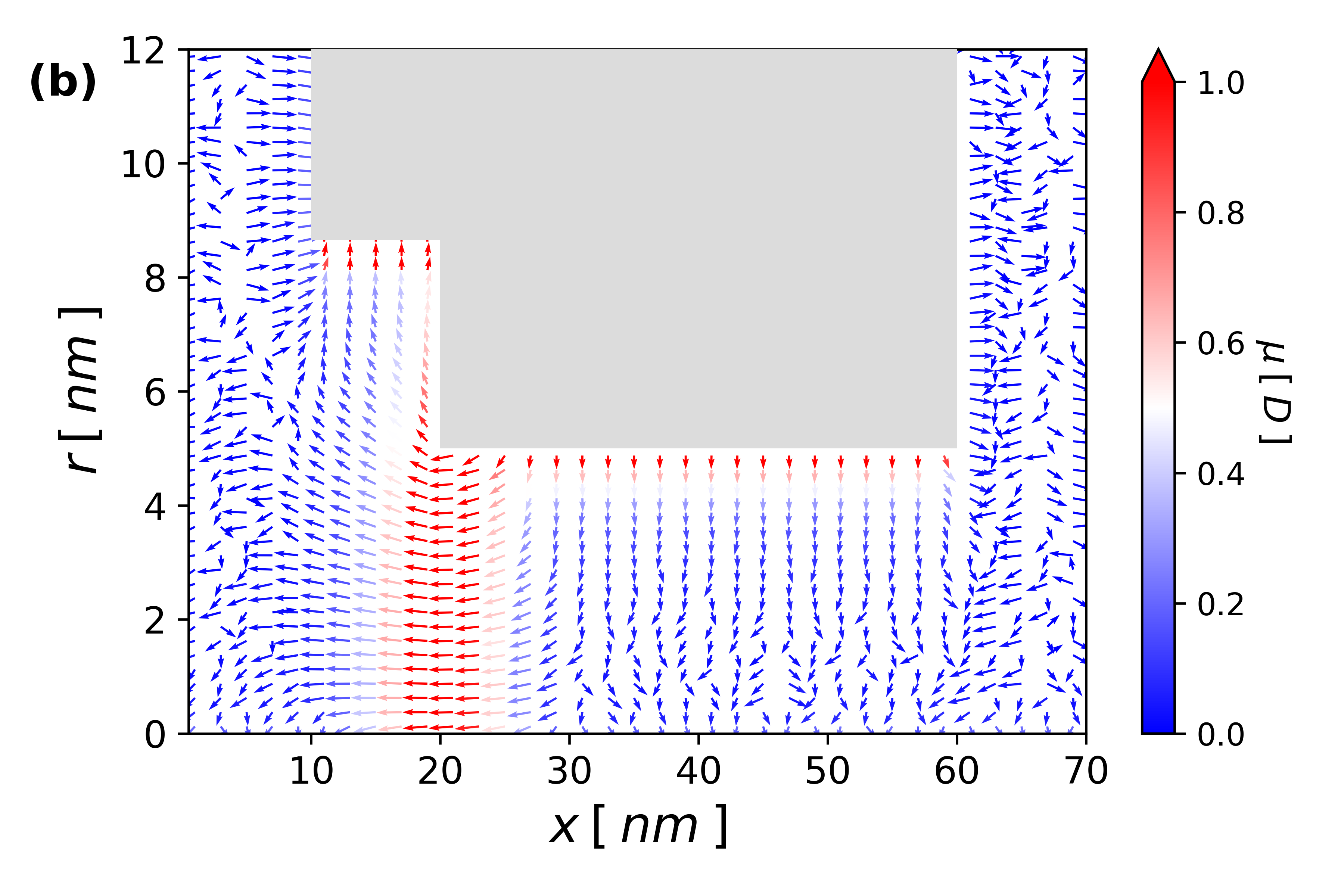}
\caption{\textcolor{black}{A 2D slice showing the dipole moment of water when the reservoir concentration is 0.05M for both the active (a) and inactive (b) pore. There is strong alignment next to the charged wall in both cases, but the peak along $x$ occurs at the respective depletion regions. Results for the 0.17 M case are shown in the supporting information Figure S 5.}}
\label{fig:dipoles}
\end{figure}

The average dipole moment under negative and positive applied potentials is shown in Figure \ref{fig:dipoles}. One can appreciate a clear tendency for water molecules to align, with both the external electric field and the electric field emanating from the charged walls. The anisotropic nature of the order spans a length that is many times larger than the size of a water molecule. There is some precedent for the dipole behaviour found here in the literature\cite{galli2014dipolar,fumagalli2018anomalouslydielectric,netz2016dielectric-planar,netz2019giantdielectric}. Past studies also proposed a reduction of the dielectric constant in the direction perpendicular to the confinement, and an increase in the parallel direction. The strength of the electrostatic interactions and, therefore, the dynamics of ionic motion, could be significantly affected by the  distinctive orientation of water at the entrance of the smaller section of the pore\cite{mio2017confinedelectros}. Importantly, although the dipole moment is altered in these regions, the average number of HB per molecule remains at 3.47, which almost exactly matches the average all across the system, including the reservoir, of 3.44. 

The results in Figure \ref{fig:dipoles} allow us to make two additional, important observations. First, as seen in panel (b), the orientation of water molecules starts inside the pore, but extends well into the reservoir, for about 5nm, highlighting the need for a reservoir when this kind of system is modelled either by molecular simulations or continuum approaches \cite{Bazant2014overlimitingcurrent}. Second, the zones displaying the strongest water alignment in Figure \ref{fig:dipoles}, \textcolor{black}{naturally correlate with the regions of augmented electric field displayed in Figure \ref{fig:electric_field}, which are also the depletion regions seen in Figure \ref{fig:concentration}. }

\begin{figure*}
\hspace*{-0cm}
\includegraphics[scale=0.65]{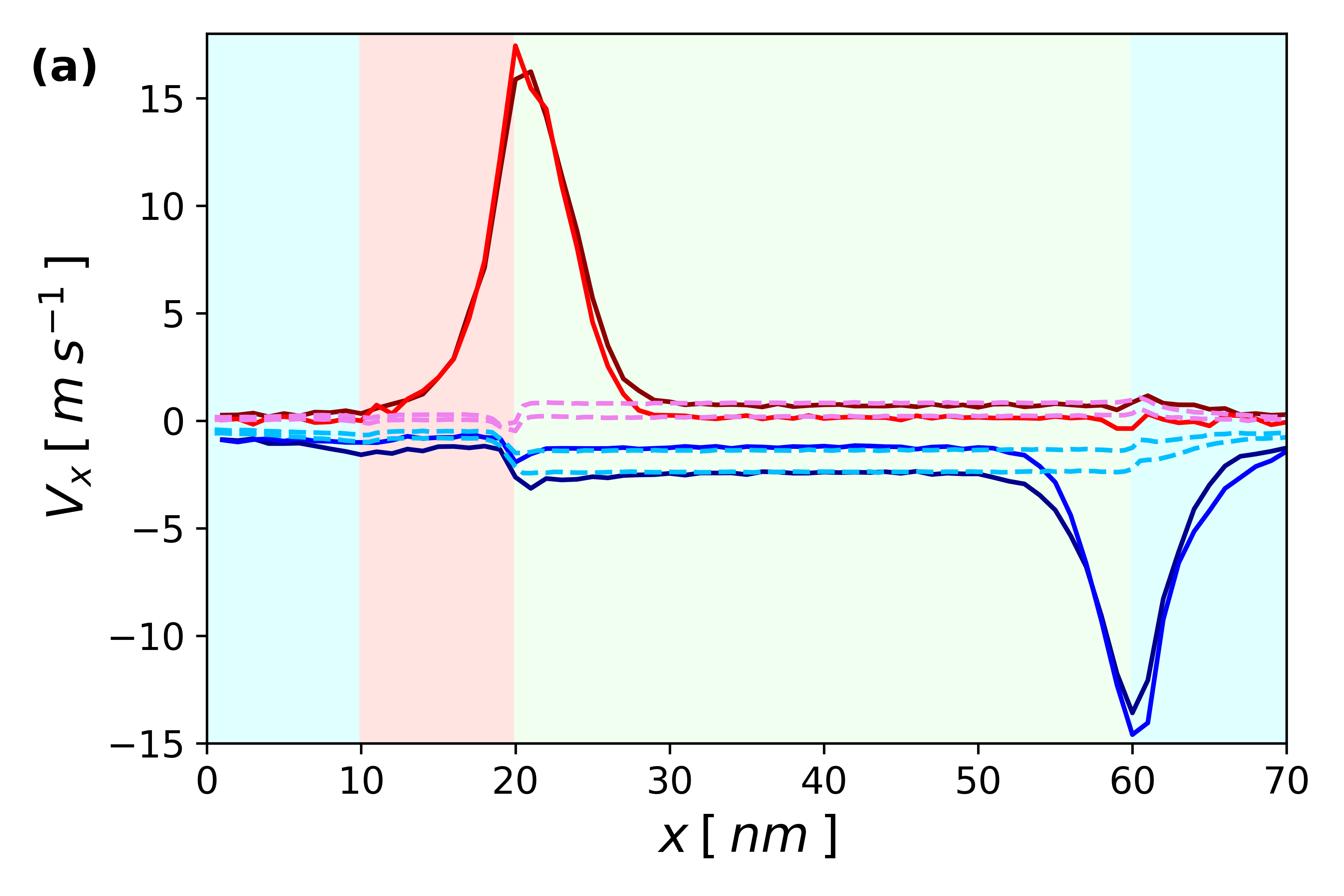}
\includegraphics[scale=0.6]{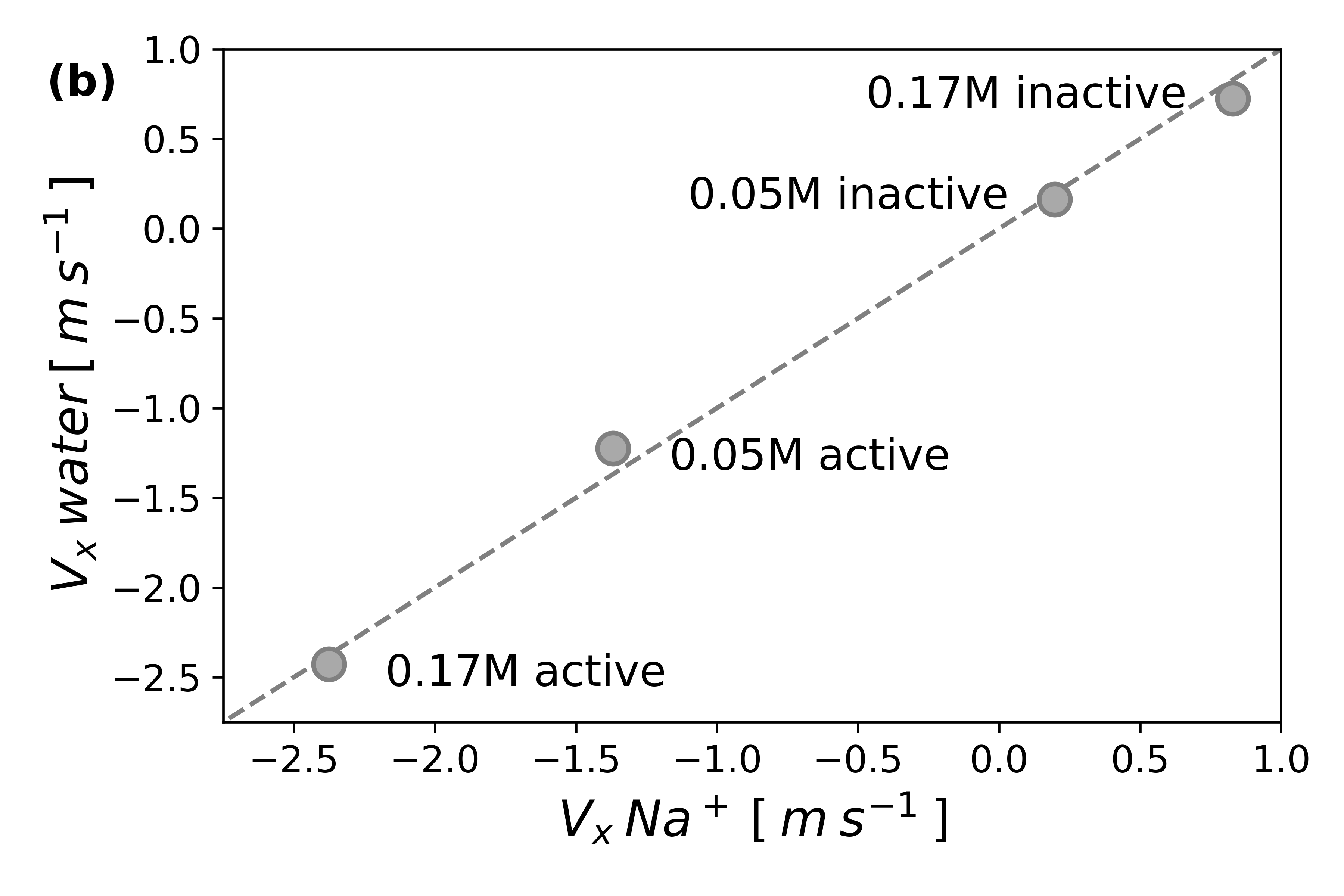}
\caption{\textcolor{black}{Water velocity averaged along the radial direction and projection onto the axis of the pore, along with the sodium result for reference, as a function of $x$ in panel (a). The full lines are sodium velocities:  blue lines are active 0.05M and 0.017M (darker), the red lines are inactive 0.05M and 0.17M(darker). The dashed lines are water velocities for inactive (pink, both concentrations) and light blue for active (both concentrations) pores.  For ease of comparison with Figure \ref{fig:velocity}, the velocities computed between $30~{\rm nm} < x < 50~{\rm nm}$ for both sodium and water are shown in panel (b). There is no lag in the sodium motion relative to the underlying solvent within the narrow pore. The complete 2D water velocity information for a reservoir concentration of 0.05M is given in the supporting information Figure S 6.}}
\label{fig:watervel}
\end{figure*}

\textcolor{black}{For completeness, in Figure \ref{fig:watervel}, we show how the water velocity developed in the radial direction is related to the sodium velocity that drives it. From this Figure, one can appreciate a clear correlation between the two velocities. In fact, away from the depletion region, water velocity exactly matches the sodium velocity. Figure \ref{fig:watervel} (b) shows how the proportionality is maintained even at different electrolyte concentrations and applied biases, for the region $(30~{\rm nm} < x < 50~{\rm nm})$. These data are consistent with previous reports for pores of smaller sizes, which have noted the concurrence of ionic current and the underlying electro-osmotic flow \cite{Bocquet2013osmotic-diode,mouterde2019angstromscale-water-nature,reviewers_paper}. This behaviour can be explained by the considerable influence that one of the ionic species (sodium in our case) has over the transport across the channel.}

\begin{figure}
\hspace*{-0cm}
\includegraphics[scale=0.25]{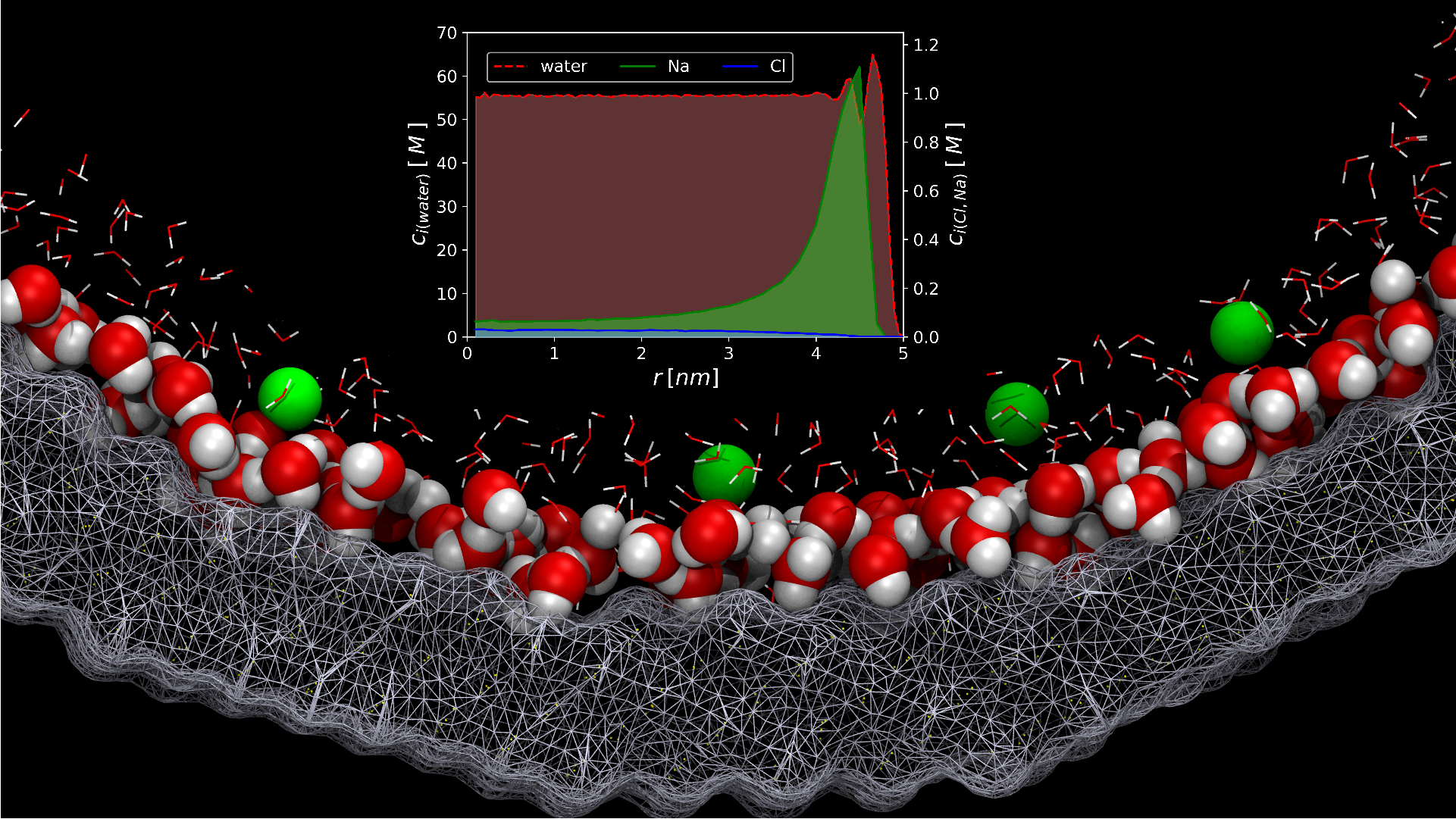}
\caption{Water mediating the contact between the ions and the walls of the pore. The radial concentration profile for water and ions in the inset reveals that sodium ions sit precisely between the first and the second layer of the water's oxygen atoms. \textcolor{black}{These results are for the 0.05M reservoir and active state simulation.}}
\label{fig:water_inbetween}
\end{figure}

\textcolor{black}{Beyond the dynamics (via coupling of sodium and solvent velocity) and complex electrostatic effects (e.g. the non-trivial dipole alignment), there is} another way in which the often overlooked presence of water is manifest in all the interactions taking place within the system. Figure \ref{fig:water_inbetween} shows the immediate vicinity of the charged wall. The sodium ions belonging to the so-called "stern layer" or "stagnant layer" are not in direct contact with the charged particles of the walls of the pore, but instead are located precisely between the first and the second layers of the water hydrating the surface. \textcolor{black}{In addition to the 'static' ion distribution, the non-zero water velocity that arises from surface roughness drags the ions in the stern layer along with it, thus impacting the dynamics of the system.} The fact that water mediates interfacial interactions certainly introduces more complexity to the already intricate structure and dynamics of the electrical double layer. 

\section{Summary}\label{sec:summary}
The results of equilibrium and non-equilibrium molecular dynamics simulations of a Janus pore in the active and inactive states are in agreement with a number of experimentally determined observables from asymmetric membranes, serving to underscore the validity of the models adopted here. 

\textcolor{black}{To electric current and rectification reported in this work can be explained in terms of an electric-field leakage model. The 'leaking' electric field arises from the applied electric field, and rearranges ions to expose wall charges. The crossing of these depletion regions at the entrance of the narrow pore, along the direction of the dominant (current carrying) sodium ions, is proposed to occur through the induced electric field. This electric field "leaks" into the reservoir for the activate case, and pulls in the cations to increase conduction. In the inactive case, the 'leakage' of the electric field does not reach into the reservoir, and instead ends in the positively charged chamber, where the sodium concentration is very small, resulting in a low output. The addition of electrolytes to the bulk (for both directions of applied electric field) should increase sodium availability, thereby enhancing electric throughput. More specifically, the enhancement depends on the increase of the local concentration in the narrow pore entrance, which depends on the spatial evolution of the double layer. Hence, to increase ICR, measures could be taken to limit sodium concentration at the narrow/wide pore junction at $x=10$ nm. Some possible design solutions could rely on extending the wider section of the membrane, or increasing its positive wall charge density.  }

Our comprehensive characterization of water structure and its influence over the membranes' performance indicates that water dipoles align not only with the externally applied electric field, but also with the field that originates from the charged walls. More importantly, water orientation is found to be highly anisotropic, with a larger average dipole moment that coincides with the ionic depletion regions. Water molecules not only align their dipoles towards the charged walls, but also in a way in which dipole-induced interactions dominate the physics at the surface of the pore. Ions, in contrast, occupy the space between the first and the second layers of water, instead of being in direct contact with the charged particles of the walls. 

The results of this work confirm that ICR is not limited to pore radii below the Debye length. \textcolor{black}{We find that the radial ion distribution in the central portion of the membrane is not too sensitive to the electrolyte concentration in the reservoir.} These findings suggest that a wider parameter space exists for which significant ICR could be obtained, thereby opening membrane design possibilities that might more easily amenable to existing fabrication processes.

\section{Model and Methods}\label{sec:formulation} 

\begin{figure}
\hspace*{-0cm}
\includegraphics[scale=0.3]{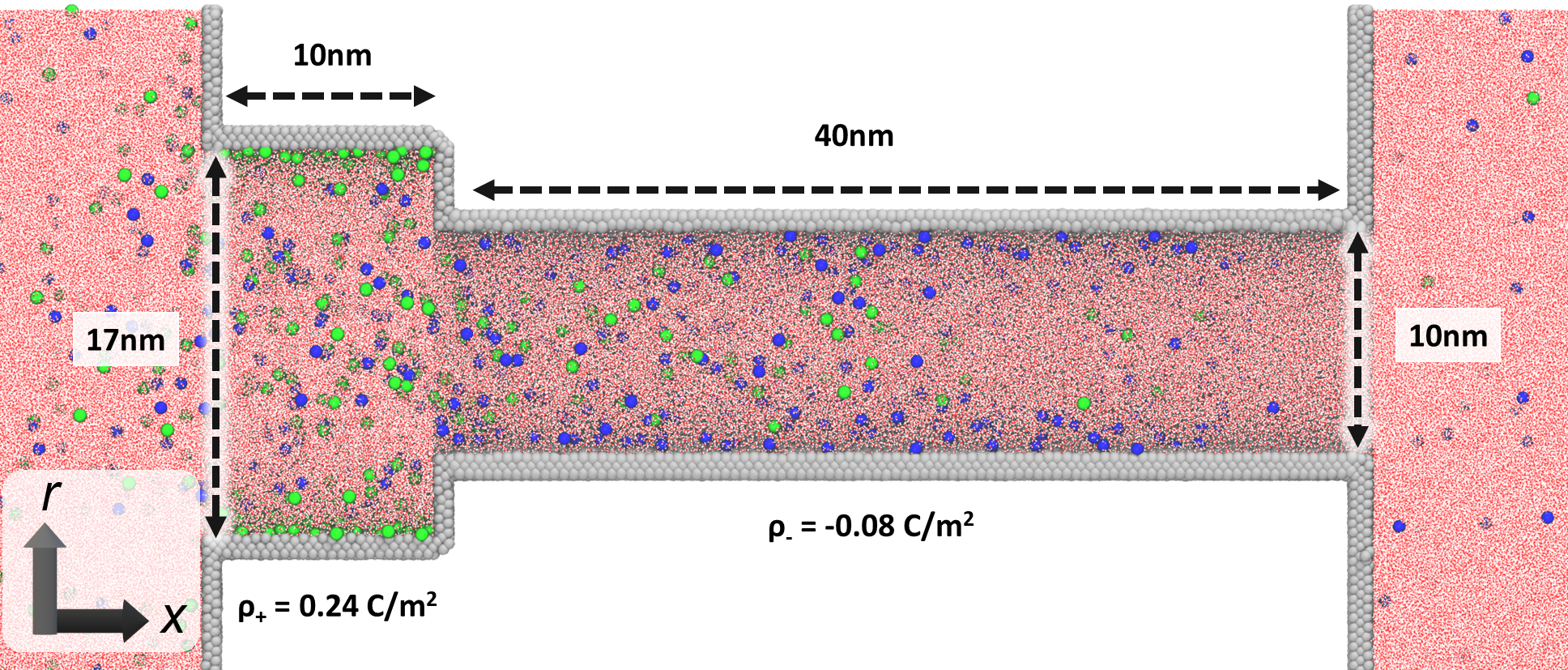}
\caption{\textcolor{black}{A 2D slice of the  model Janus membrane used in our molecular dynamics simulations. The grey spheres are the fixed walls, pink represents water molecules, blue are sodium ions, and green are chloride ions. Only the horizontal walls are charged.}}
\label{fig:sketch}
\end{figure}

In the experiments by Zhang \latin{et al.}\citep{zhang2017ultrathin}, the active part of the Janus pore consisted of two contiguous cylindrical sections. The narrow section had a surface charge of -0.08 ${\rm C}/{\rm m}^2$, while the wider section had a surface charge of +0.24 ${\rm C}/{\rm m}^2$. We maintained that designed in our simulations but re-scaling the length of the pore to 10 nm at the wider section and 40 nm at the narrower part as to preserve the original length and diameter ratio between the oppositely charged sections \textcolor{black}{and still be numerically feasible.} In our model a 'reservoir' with dimensions 20 nm $x$ 28.5 nm $x$ 28.5 nm is connected to both sides of the pore. A schematic representation of the system is shown in Figure \ref{fig:sketch}. This Janus membrane is operated with a NaCl 1:1 electrolyte solution.
The molecular simulations were conducted under the assumption that the transport properties in the real membrane consists of the sum of the transport of many non-interacting independent pores working in parallel. Under this supposition, the behaviour of the membrane is expected to be recovered from the simulation of a single pore. This is justified by the negligible change in conductance for different pore arrangement densities at ionic concentrations similar to ours in the experiments by Zhu \latin{et al.}\cite{zhu2018JANus-rectification_factor2}.

For simulations, the nanopore was carved from an FCC \textcolor{black}{structure with a lattice constant of 0.407 nm.} The particles of the walls are modelled with Lennard-Jones parameters sigma=0.32nm and epsilon=0.65 kJ/mol. The small section of the channel is composed of 41691 particles, each carrying a charge of Q=-0.01504 a.u. to achieve the desired surface charge. Similarly, the large section of the channel contains 17921 particles, each carrying a charge of 0.04464 in atomic units. The channel is embedded in a box with dimensions provided in Figure 1, and is filled with approximately 705000 water molecules. Using pre-equilibrated NaCl solutions, two simulations are performed, with 2234 and 4393 ions, in an overall electro-neutral simulation box. For the electrolyte solution, we use the Dang parameters~\cite{DandIons} for the ions and SPC/E~\cite{spce} for the water.

All the MD simulations were performed using GROMACS~\cite{berendsen1995gromacs} in the NVT ensemble with periodic boundary conditions. Long-range electrostatic forces were calculated using the particle-mesh Ewald method, with a grid density of 0.16 nm and a real space cut-off at 1.2 nm. The temperature of all simulations was kept constant at 300K using a velocity-rescaling algorithm \textcolor{black}{coupled every 0.1 ps with no centre of mass motion removal.} The positions of all the particles of the walls were restrained using harmonic potentials \textcolor{black}{of $9000 \, kJ\,.mol^{-1}.nm^{-2}$} over the entire course of the simulations. A uniform electrostatic field of (+/-) 0.0333 V/nm, \textcolor{black}{  corresponding to activate and inactive, states respectively,} was applied in the axial direction to produce the ionic current.

\textcolor{black}{The entire system was first equilibrated without external bias for a period of 30 ns. After that, a production simulation was run for another 30 ns to generate the reference equilibrium state of the pore. Starting from the equilibrium state, intermediate simulations were conducted at each different potential for a period of 30 ns until steady state was reached. The simulations were continued for an additional 30ns period for the final data collection. The steady state of the final runs was determined by inspecting the ionic current evolution over the course of the simulations (see Supporting Information Figure S 1).}

\begin{suppinfo}

The ionic current generated in the activated and inactivated pore as a function of time (Figure S 1). For the 0.17 M reservoir the  2D salt concentration distribution for all three biases (Figure S 2); Radially averaged electric field in the active and inactive states (Figure S 3); 2D Water dipole moment (Figure S 5); and streamlines of water motion (Figure S 6). The ion streamlines are shown at a reservoir salt concentration of 0.05 M (Figure S 4).

\end{suppinfo}

\section{Acknowledgments}\label{sec:Acknowledgments}
This work was supported as part of the Advanced Materials for Energy-Water Systems (AMEWS) Center, an Energy Frontier Research Center funded by the U.S. Department of Energy, Office of Science, Basic Energy Sciences. 

\bibliography{revised_arxiv}

\pagebreak
\begin{center}
\textbf{\large Supplemental Materials: Ionic transport in electrostatic Janus Membranes. An explicit solvent molecular dynamic simulation.}
\end{center}

\setcounter{equation}{0}
\setcounter{figure}{0}
\setcounter{table}{0}
\setcounter{page}{1}
\makeatletter
\renewcommand{\theequation}{S\arabic{equation}}
\renewcommand{\thefigure}{S\arabic{figure}}
\renewcommand{\bibnumfmt}[1]{[S#1]}
\renewcommand{\citenumfont}[1]{S#1}

\begin{figure}
\hspace*{-0cm}
\includegraphics[scale=1]{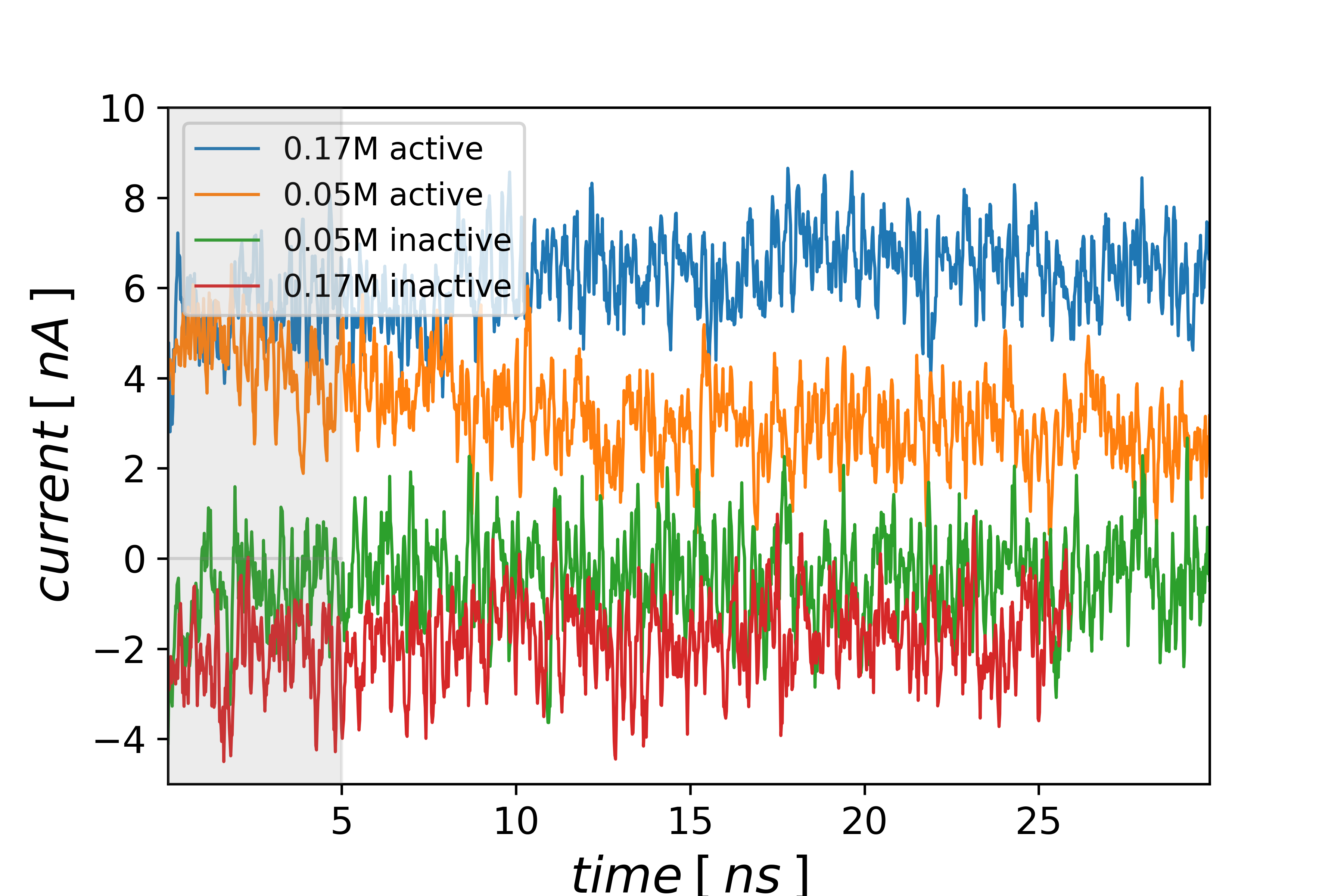}
\caption{The ionic current versus time is shown for the active and inactive modes at both 0.05M and 0.17M reservoir salt concentrations. There are high frequency fluctuations but the current has reached steady state over the reported time period. Hence it is expected that the simulations have converged.}
\label{fig:ion_current_vs_time}
\end{figure}

\begin{figure*}
\hspace*{-0cm}
\includegraphics[scale=0.75]{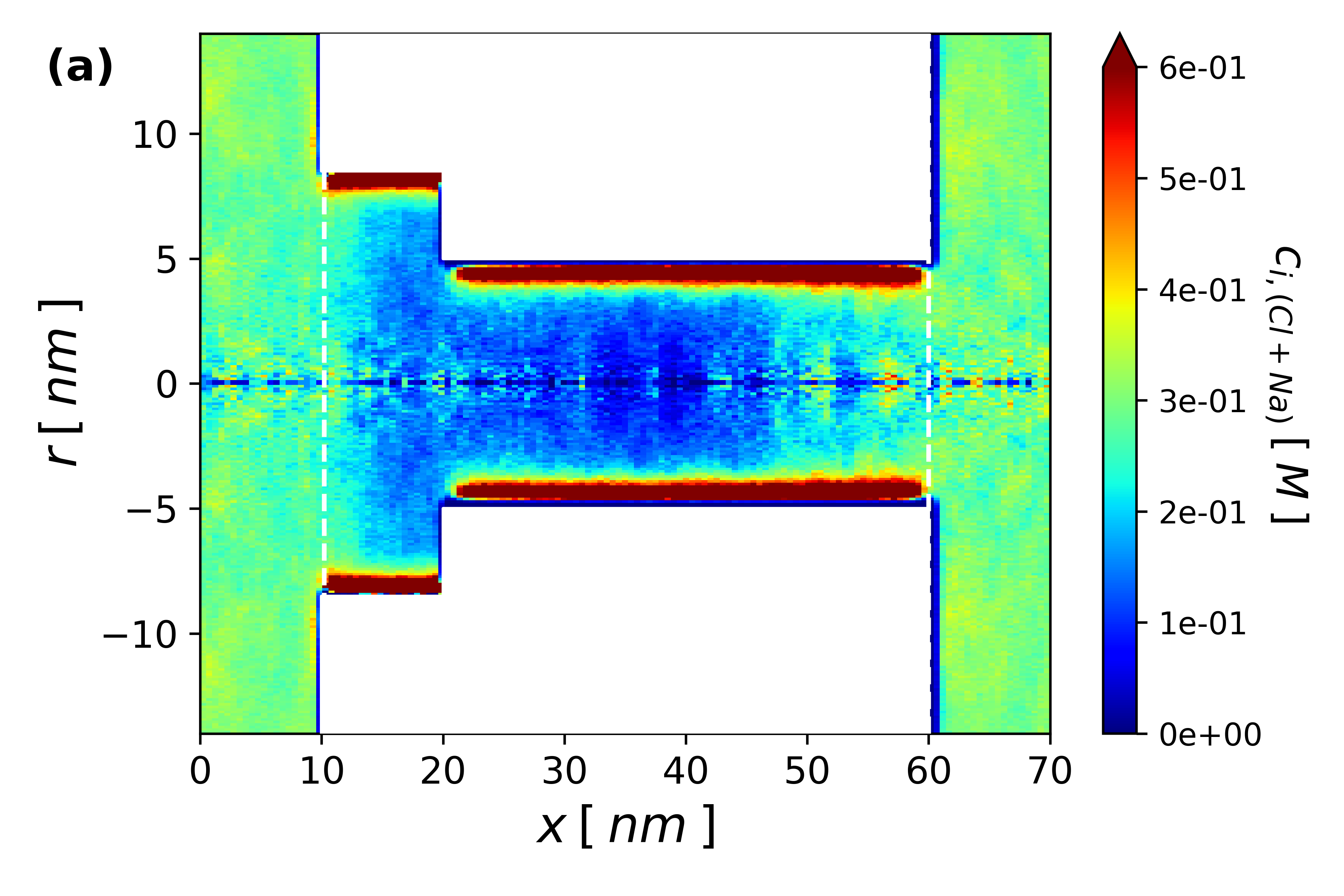}
\includegraphics[scale=0.75]{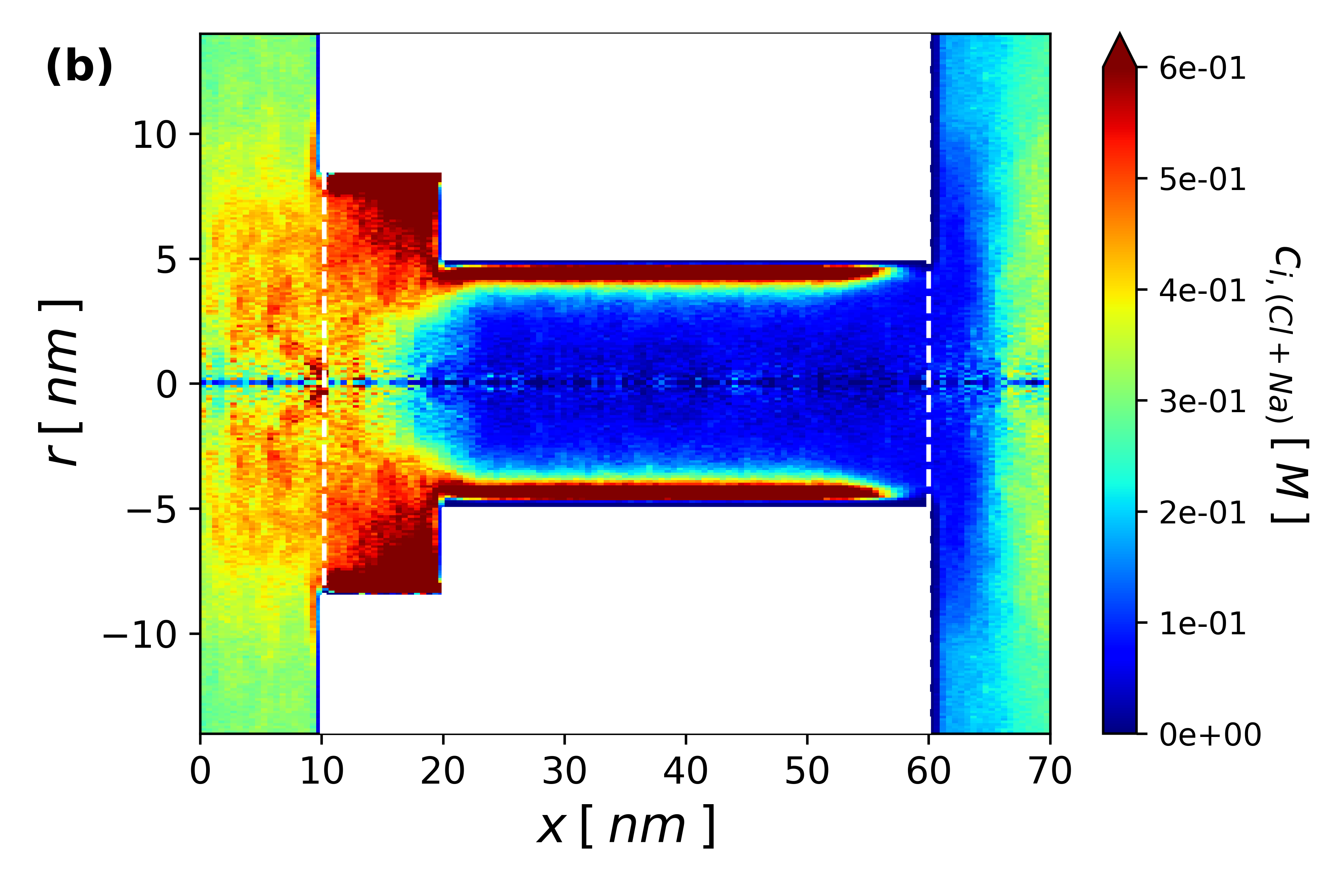}
\includegraphics[scale=0.75]{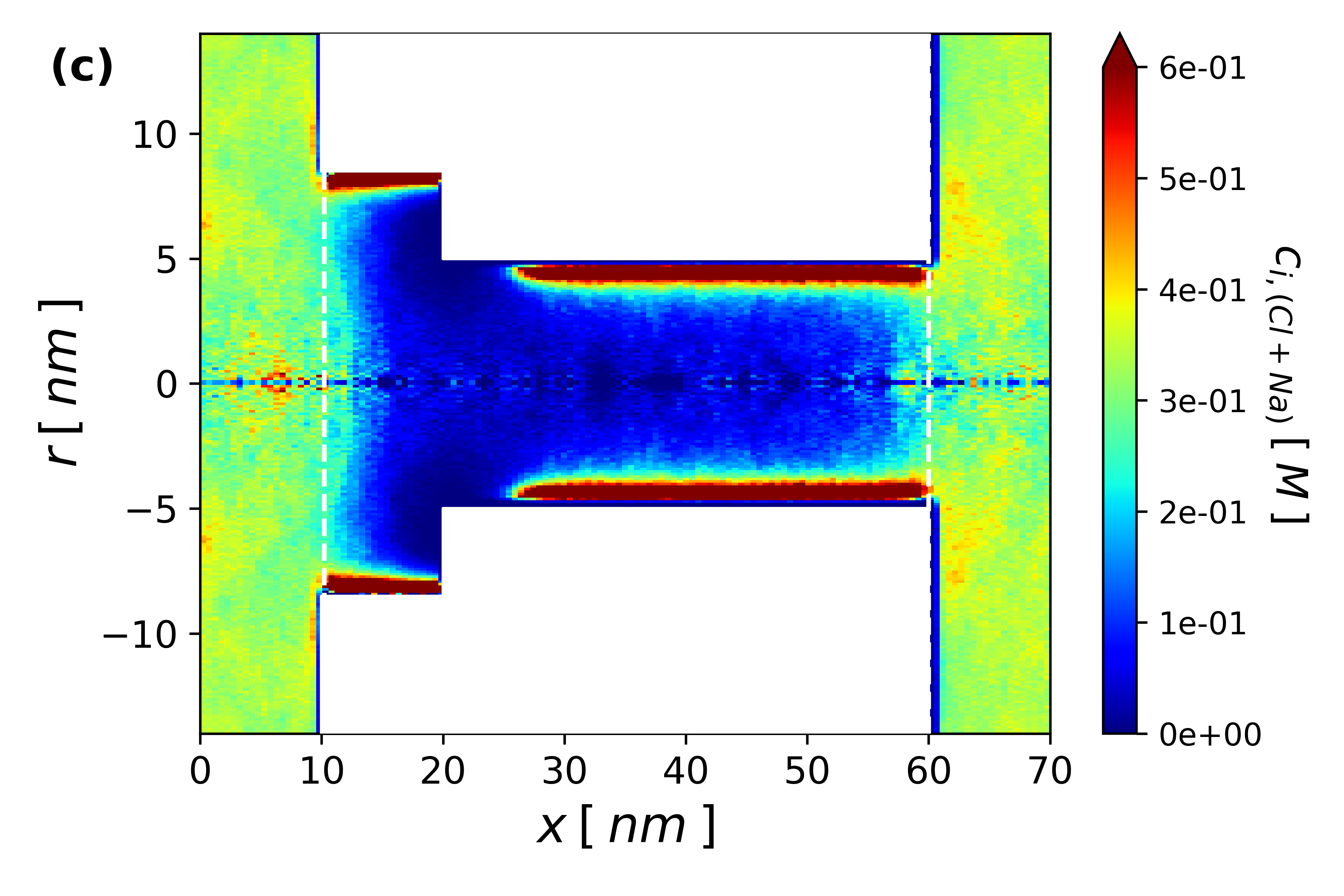}
\caption{2D Ionic (Na+Cl) concentration distribution for a reservoir concentration of 0.17M. The top is the equilibrium state at 0V, the middle is the activated state while the bottom is the inactivated case.}
\label{fig:concentration_SI}
\end{figure*}

\begin{figure*}
\hspace*{-0cm}
 \includegraphics[scale=0.7]{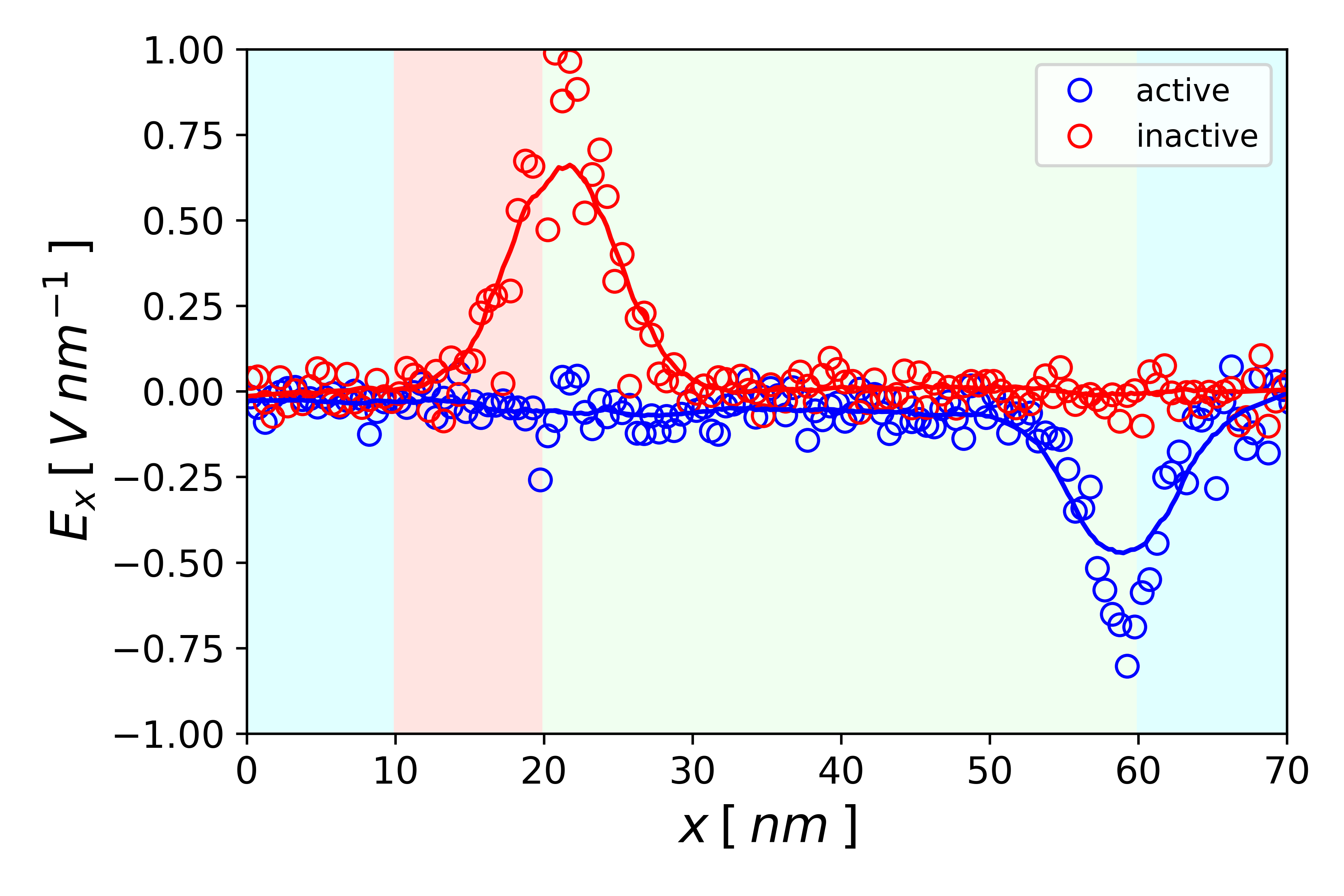}
\caption{The radially averaged local electric field along the $x$ direction is shown as a function of the axial coordinate $x$ for the active and the inactivate states at a reservoir salt concentration of 0.17M. These results are nearly identical to those in the main figure 4 performed at 0.05M reservoir concentration.}
\label{fig:electric_field_SI}
\end{figure*}

\begin{figure}
\hspace*{-0cm}
\includegraphics[scale=0.65]{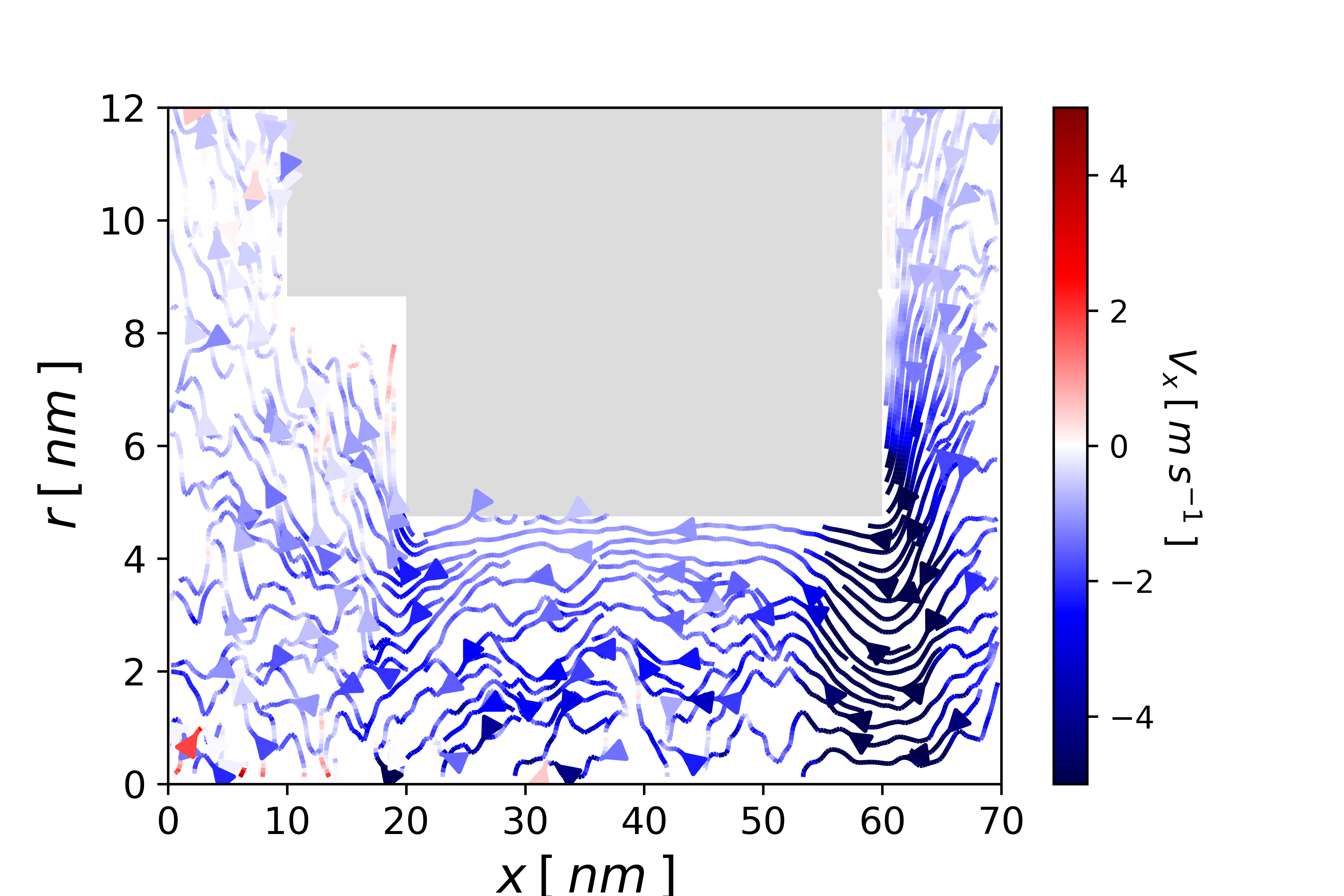}
\includegraphics[scale=0.65]{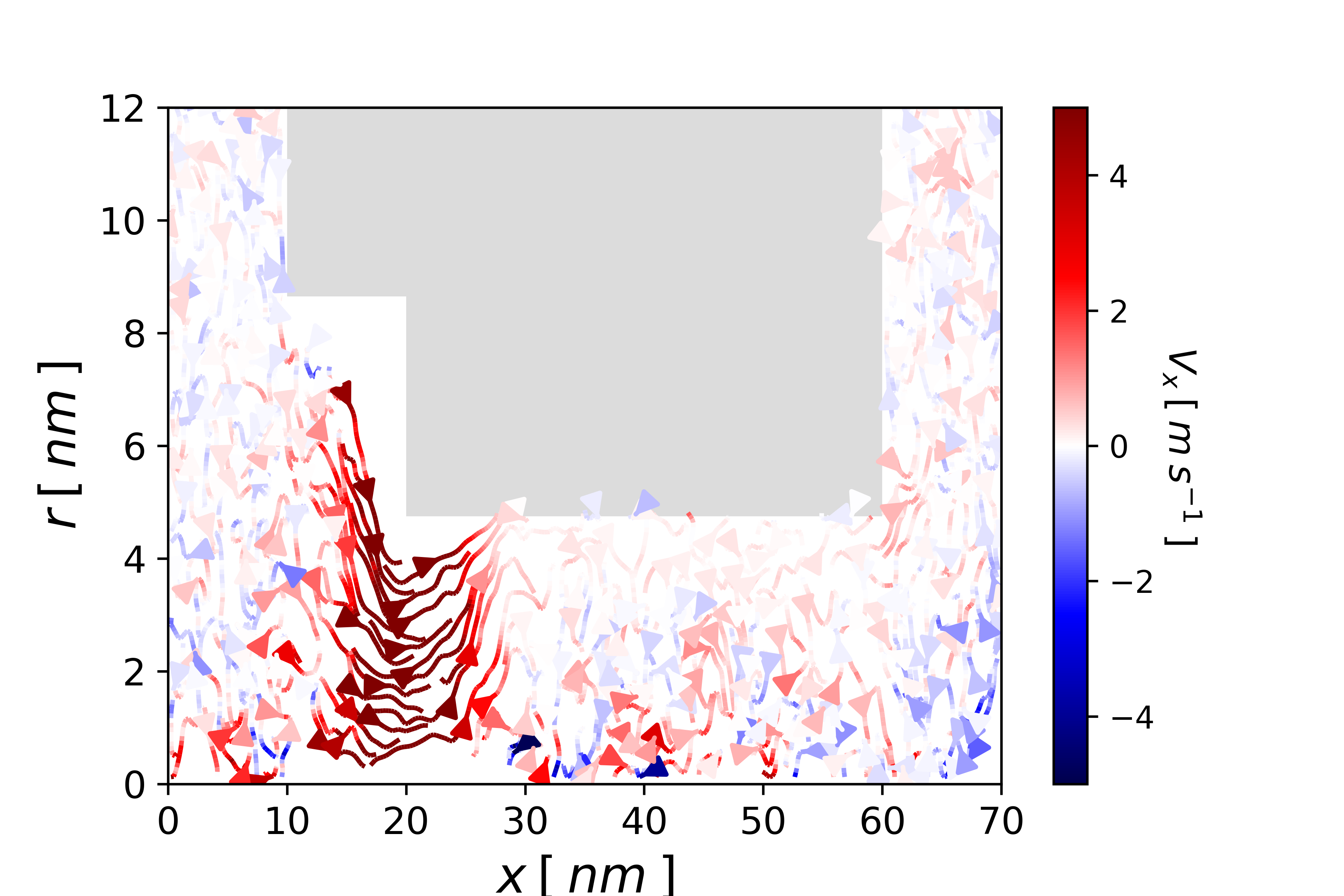}
\caption{Streamlines of the ion velocity are shown at 0.05M reservoir concentration for the active (a) and inactive (b) states. The radially averaged result is shown in figure 5.}
\label{fig:velocity_SI}
\end{figure}

\begin{figure}
\hspace*{-0cm}
\includegraphics[scale=0.65]{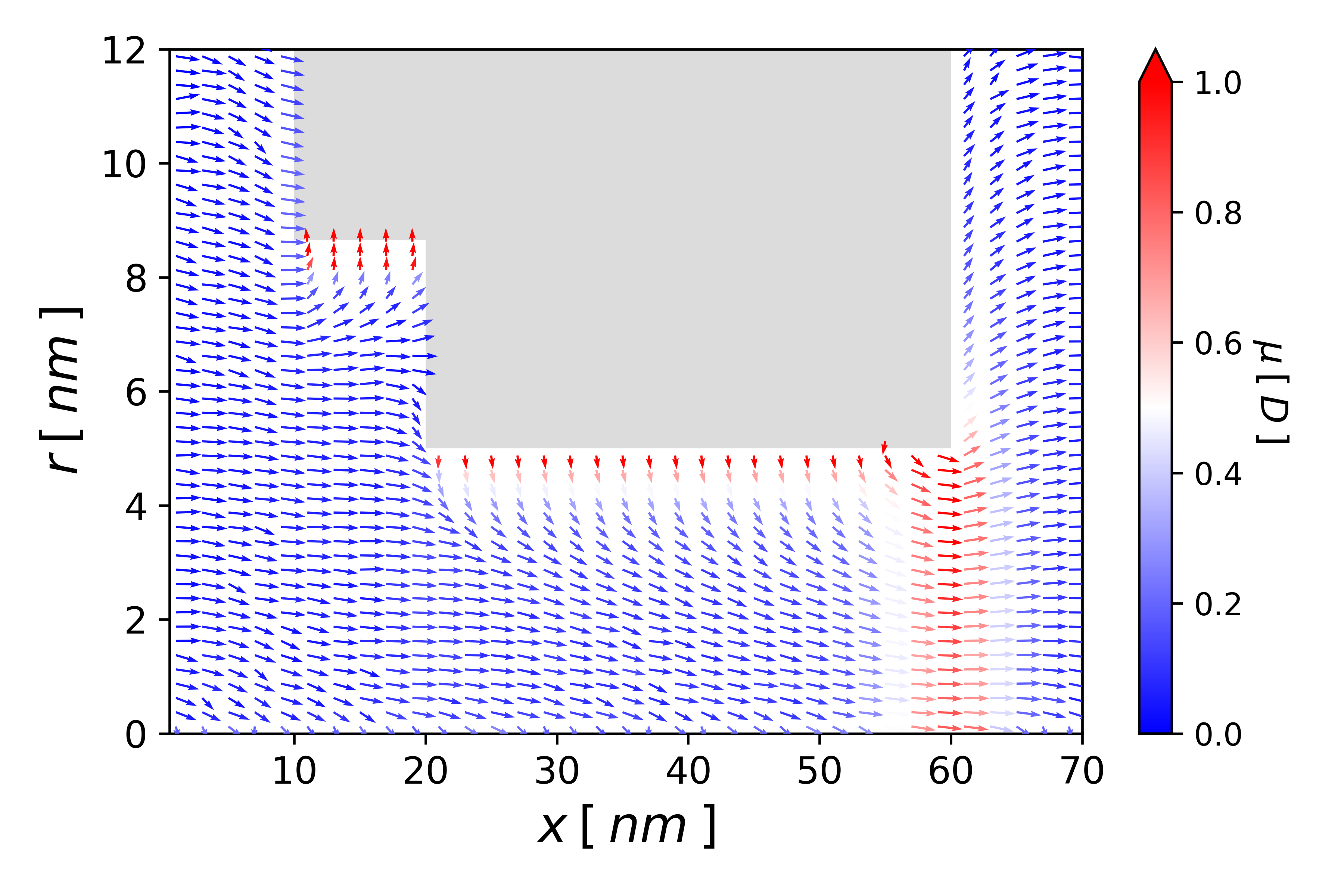}
\includegraphics[scale=0.65]{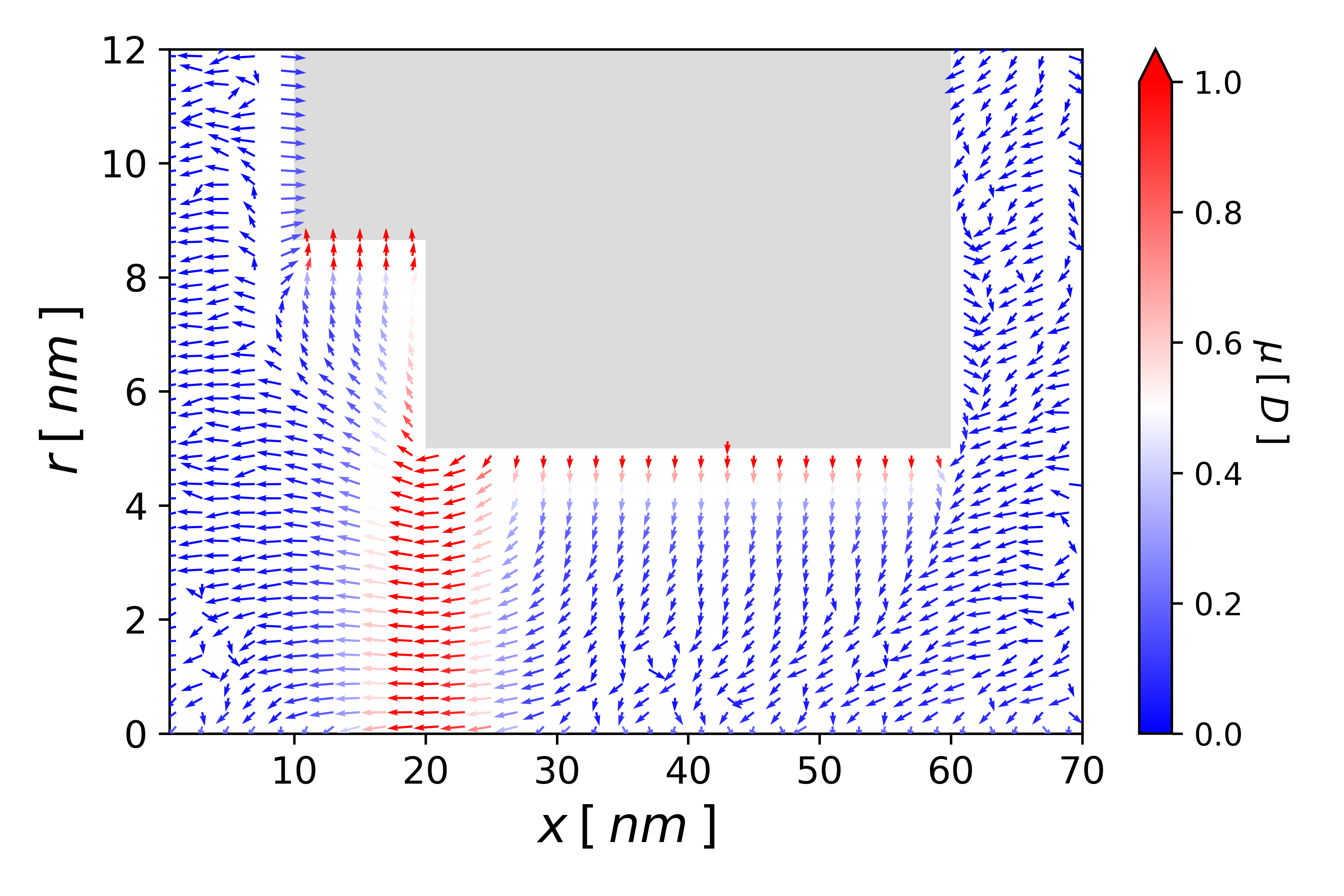}
\caption{At 0.17M reservoir concentration the dipole moment of water is shown for both the active (a) and inactive (b) pore. This results do not strongly differ from the 0.05M result shown in main figure 7.}
\label{fig:dipoles_SI}
\end{figure}

\begin{figure}
\hspace*{-0cm}
\includegraphics[scale=0.65]{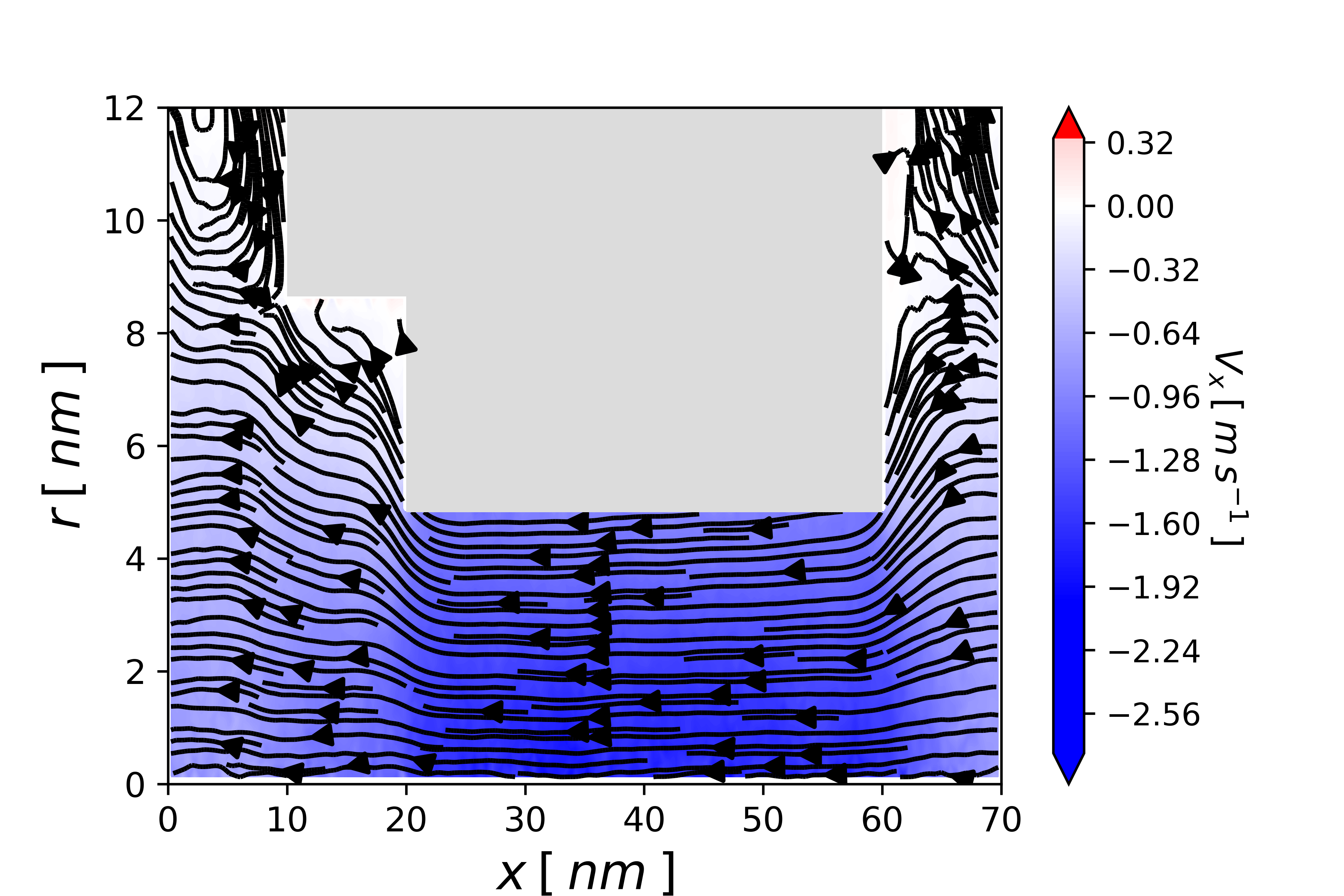}
\includegraphics[scale=0.65]{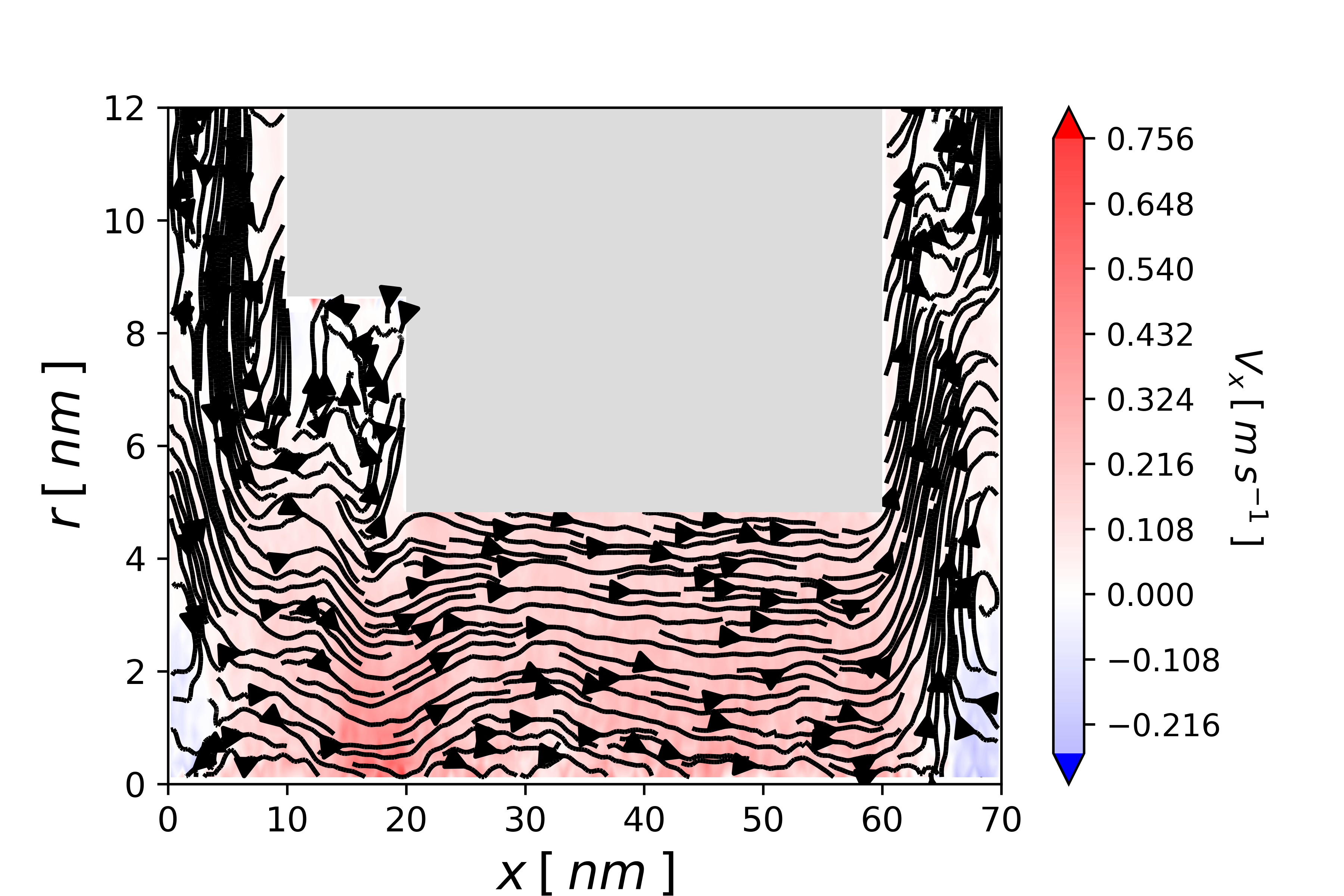}
\caption{Water streamlines for the activated (a) and inactivated (b) pores are shown for calculations performed at 0.05M. The radially averaged result is shown in figure 8 (a).}
\label{fig:water_vel_SI}
\end{figure}

\end{document}